\documentclass[aps,prb,reprint,showpacs,superscriptaddress,floatfix]{revtex4-1}
\usepackage[utf8]{inputenc}
\usepackage{amsfonts}
\usepackage[intlimits]{amsmath}
\usepackage{bbm,xcolor}
\usepackage{stmaryrd}
\usepackage{graphicx}
\usepackage[plain]{fancyref}
\usepackage{hyperref}
\usepackage{xspace}
\usepackage{multirow}

\newcommand{\Eq}[1]{Eq.~(\ref{#1})}
\newcommand{\Eqs}[1]{Eqs.~(\ref{#1})}
\newcommand{\id}{\mathbbm{1}}

\def\bra#1{\ensuremath{\langle{#1}\vert}}
\def\ket#1{\ensuremath{\vert{#1}\rangle}}
\def\bracket#1#2{\ensuremath{\langle{#1}\mkern1.2mu\vert\mkern1.2mu{#2}\rangle}}
\def\expect#1{\ensuremath{\langle{#1}\rangle}}
\def\abs#1{\ensuremath{\left|{#1}\right|}}
\def\norm#1{\mathinner{\lVert#1\rVert}}
\def\figref#1{\figurename~\ref{#1}}

\def\vec#1{{\boldsymbol{#1}}}
\newcommand{\hc}[0]{{\rm h.c.}}

\newcommand{\chebyps}{Chebyshev polynomials\xspace}
\newcommand{\cheby}{Chebyshev\xspace}
\newcommand{\obroadening}{optimal broadening\xspace}
\newcommand{\qph}{\quad \phantom{.}}
\newcommand{\qqph}{\qquad \phantom{.}}
\newcommand{\Hshift}{H'}

\newcommand{\wshift}{\omega'}
\newcommand{\eg}{e.\,g.\ }

\newcommand{\ham}{Hamiltonian\xspace}
\newcommand{\Bop}{\hat{\mathcal{B}}}
\newcommand{\Cop}{\hat{\mathcal{C}}}

\newcommand{\WA}{{W_{\! \! \mathcal{A}}}}
\newcommand{\Wast}{W_{\! \ast}} %{W_{\rm eff}}

\newcommand{\etaN}{\eta_N}
\newcommand{\epsilont}{\epsilon_{\rm t}} %{\epsilon_{\rm s}}
\newcommand{\dtrunc}{d_K} %{d_{\rm p}}
\newcommand{\ntrunc}{n_S} %{n_{\rm p}}

\newcommand{\etrunc}{\varepsilon_P} %{\varepsilon_{\rm P}}
\newcommand{\trunc}{{\rm tr}} %{{\rm p}}

\newcommand{\sweep}{{\rm sweep}}
\newcommand{\Wfull}{W}
\newcommand{\CheMPS}{CheMPS\xspace}

\newcommand{\ED}{{\rm ED}}
\newcommand{\Nmax}{N_{\rm max}}
\newcommand{\Ch}{{\rm Ch}}

\renewcommand{\figurename}{Fig.}

\begin{document}

\title{Chebyshev matrix product state approach for spectral functions}

\author{Andreas \surname{Holzner}}
\author{Andreas \surname{Weichselbaum}}
\affiliation{Physics Department, Arnold Sommerfeld Center for Theoretical Physics, and Center for NanoScience,
  Ludwig-Maximilians-Universität München, D-80333 München, Germany}
\author{Ian P. \surname{McCulloch}}
\affiliation{School of Physical Sciences, University of Queensland, Brisbane, Queensland 4072, Australia}
\author{Ulrich \surname{Schollwöck}}
\author{Jan \surname{von Delft}}
\affiliation{Physics Department, Arnold Sommerfeld Center for Theoretical Physics, and Center for NanoScience,
  Ludwig-Maximilians-Universität München, D-80333 München, Germany}

\date[Date: ]{\today}

\begin{abstract}
  We show that recursively generated \cheby expansions offer
  numerically efficient representations for calculating
  zero-temperature spectral functions of one-dimensional lattice
  models using matrix product state (MPS) methods. The main features
  of this Chebychev matrix product state (\CheMPS) approach are: (i)
  it achieves \emph{uniform} resolution over the spectral function's
  entire spectral width; (ii) it can exploit the fact that the latter
  can be much \emph{smaller} than the model's many-body bandwidth;
  (iii) it offers a well-controlled \emph{broadening} scheme that
  allows finite-size effects to be either resolved or smeared out, as
  desired; (iv) it is based on using MPS tools to
  \emph{recursively} calculate a succession of Chebychev vectors
  $|t_n \rangle$,
  (v) whose \emph{entanglement entropies} were found to remain
  bounded with increasing recursion order $n$ for all cases
  analyzed here; (vi) it \emph{distributes} the \emph{total
    entanglement entropy} that accumulates with increasing $n$
  over the set of \cheby vectors $|t_n \rangle$, which need not be
  combined into a single vector. In this way, the growth in
  entanglement entropy that usually limits density matrix
  renormalization group (DMRG) approaches is packaged into
  conveniently manageable units.
  We present zero-temperature \CheMPS results for the
  structure factor of spin-$\frac{1}{2}$ antiferromagnetic
  Heisenberg chains and perform a detailed finite-size
  analysis. Making comparisons to three benchmark methods, we find
  that CheMPS (1) yields results comparable in quality to those of
  correction vector DMRG, at dramatically reduced numerical cost;
  (2) agrees well with Bethe Ansatz results for an infinite system,
  within the limitations expected for numerics on finite systems;
  (3) can also be applied in the time domain, where it has potential
  to serve as a viable alternative to time-dependent DMRG (in
  particular at finite temperatures). Finally, we present a detailed
  error analysis of \CheMPS for the case of the noninteracting
  resonant level model.
\end{abstract}


\maketitle

\section{Introduction}
\label{sec:intro}

Consider a one-dimensional lattice model amenable to treatment by
the density matrix renormalization group
(DMRG),\cite{white:dmrg1,white:dmrg2,Schollwock2005,Schollwoeck2010}
with Hamiltonian $\hat H$, ground state $|0 \rangle$ and ground
state energy $E_0$.  This paper is concerned with zero-temperature
spectral functions of the form
\begin{eqnarray}
  \label{eq:spectral-frequency}
  \mathcal{A}^{\mathcal{BC}}(\omega) = \bra{0} \Bop \, 
  \delta(\omega - \hat H + E_0) \, \Cop \ket{0}  \; , 
\end{eqnarray}
which represents the Fourier transform $\int \frac{d t}{2 \pi}
e^{ i \omega t}  G^{\mathcal{BC}} (t) $
of the correlator
\begin{eqnarray}
  \label{eq:GAB}
  G^{\mathcal{BC}} (t) 
  = \langle 0 | \Bop(t)\Cop(0) | 0 \rangle \; . 
\end{eqnarray}
One possible framework for calculating such spectral functions is to
expand them in terms of Chebychev polynomials, as advocated in
Ref.~\onlinecite{Weisse2006}. Such a \cheby expansion offers precise
and convenient control of the accuracy and resolution with which a
spectral function is to be computed.  
This is very useful, particularly when broadening the spectral
function of a length-$L$ system, which exhibits finite-size subpeaks
with spacing $\omega_L \sim 1/L$, in order to mimick that of an
infinite system: if the latter has structures (e.g.\ sharp or
diverging peaks) which are not yet properly resolved at the scale
$\omega_L$, the broadened version of the finite-size spectral function
inevitably bears $L$-dependent errors in the vicinity of these
structures. Hence, when calculating the finite-size version of these
structures for the length-$L$ system, there is no need to achieve an
accuracy beyond that of the expected $L$-dependent errors, and having
convenient control of this accuracy can significantly reduce numerical
costs.

In this paper, we show that Chebyshev expansions offer numerically
efficient representations for calculating spectral functions using
matrix product state (MPS)
methods,\cite{dukelsky1998,Vidal2004,verstraete:dmrg-pbc,Verstraete2004b,McCulloch2007,Schollwoeck2010}
with numerical costs that compare favorably to those of other
established DMRG-based approaches. In particular, the Chebychev MPS
approach presented here, to be called \CheMPS, allows the
abovementioned control of accuracy and resolution to be imported into
the DMRG/MPS arena.

The historically first approach for calculating spectral functions
with DMRG is the continued-fraction expansion.\cite{Hallberg1995}
While this method requires only modest numerical resources, it is
limited to low frequencies and it is difficult to produce reliable
results with it in the case of continua (however, algorithmic
improvements were reported recently\footnote{It was shown very
  recently\cite{DargelHoneckerPetersNoackPruschke2010} that the
  performance of the continued-fraction approach can be substantially
  improved by iteratively calculating its expansion coefficients using
  an adaptive Lanczos-vector method.}). At present, the most accurate,
but also most time-consuming approaches are: (i) the correction vector
(CV)
method,\cite{RamaseshaPatiKrishnamurthyShuaiBredas1997,Kuhner1999,Jeckelmann2002} and
(ii) time-dependent DMRG
(tDMRG),\cite{Vidal2004,White2004,DaleyKollathSchollwoeckVidal2004,Verstraete2004b,Schmitteckert2004}
in particular when combined with linear prediction
techniques.\cite{PressTeukolskyVetterlingFlannery2007,PereiraWhiteAffleck2008,WhiteAffleck2008,BarthelSchollwoeckWhite2009}
Since any new approach must measure up to their standards, let us
briefly summarize their key ideas, advantages and drawbacks.

(i) To calculate $  \mathcal{A}^{\mathcal{BC}}(\omega)$
using the CV approach, it is expressed as 
\begin{subequations}
  \label{eq:CV}
  \begin{eqnarray}
    \label{eq:CV-1}
    \mathcal{A}^{\mathcal{BC}}(\omega) =  
    \langle 0 | \Bop | \mathcal{C}\rangle_\omega \; , 
  \end{eqnarray}
  in terms of the so-called correction vector
  \begin{eqnarray}
    \label{eq:CV-2}
    | \mathcal{C} \rangle_\omega \equiv  - \lim_{\eta \to 0} 
    \frac{1}{\pi} {\rm Im} \left[  \frac{1}{\omega - \hat H + E_0 + i \eta} 
    \right] \Cop |0\rangle \; .
  \end{eqnarray}
\end{subequations}
The correction vector can be calculated (for finite broadening
parameter $\eta$) using either conventional DMRG
\cite{RamaseshaPatiKrishnamurthyShuaiBredas1997,Kuhner1999,Jeckelmann2002}
or variational matrix product state (MPS)
methods.\cite{Weichselbaum2009}
A major advantage of this
approach is that arbitrarily high spectral resolution can be achieved
by reducing $\eta$ and sampling enough frequency points. However, this
comes at considerable numerical costs: first, a separate calculation
is required for every choice of $\omega$ (though in doing so,
results for $|\mathcal{C} \rangle_\omega$'s from previous frequencies
can be incorporated); and second, the calculation of $|\mathcal{C}
\rangle_\omega$ involves an operator inversion problem that is
numerically poorly conditioned, ever more so the smaller $\eta$ is.

(ii) An alterative possibility is to use tDMRG to calculate the
time-domain correlator $G^{\mathcal{BC}}(t)$, Fourier transforming to
the frequency domain only at the very end.  To this end, one expresses
\begin{subequations}
  \label{eq:tDMRG}
  \begin{eqnarray}
    \label{eq:tDMRG-1}
    G^{\mathcal{BC}} (t)  = e^{i E_0 t} \langle 0 | \Bop  | \tilde {\mathcal{C}}\rangle_t \;  
  \end{eqnarray}
  in terms of the time-evolved state 
  \begin{eqnarray}
    \label{eq:Ct}
    | \tilde {\mathcal{C}}\rangle_t \equiv e^{-i \hat H t} \Cop |0 \rangle \; 
  \end{eqnarray}
\end{subequations}
and uses tDMRG to calculate the latter. Two attractive features of
this strategy are: first, it builds on an extensive body of
algorithmic knowledge for efficiently calculating
time-evolution;\cite{Vidal2004,White2004,DaleyKollathSchollwoeckVidal2004}
and second, a simple linear-prediction scheme
\cite{PressTeukolskyVetterlingFlannery2007,PereiraWhiteAffleck2008,WhiteAffleck2008,
  BarthelSchollwoeckWhite2009} can be used to extrapolate the
time-dependence calculated for short and intermediate time scales to
longer times, thereby improving the quality of results at low
frequency at hardly any additional numerical cost. However, obtaining
reliable results over a sufficiently large time interval can, in
itself, be numerically very expensive, since the time-evolution of the
many-body state $|\tilde{\mathcal{C}} \rangle_t$
is accompanied by a strong growth in entanglement entropy.
This unavoidably also implies a growth of tDMRG truncation errors.

Note that in both of the schemes outlined above, significant
(often heroic) amounts of numerical resources are devoted to
calculating a single state, $|\mathcal{C} \rangle_\omega$ for given
$\omega$ or $|\tilde{\mathcal{C}} \rangle_t$ for given $t$, as
accurately as possible; the overlaps or expectation values of
interest, namely $\langle 0 | \Bop |\mathcal{C}\rangle_\omega$ for
$\langle 0 | \Bop|\tilde{\mathcal{C}} \rangle_t$, are only
calculated \emph{at the end}, in a single, final step, after
$|\mathcal{C} \rangle_\omega$ or $|\tilde{\mathcal{C}} \rangle_t$
have been fully determined.  Actually, these states are calculated
so accurately that they would have been equally suitable for
calculating any other quantity (correlator or matrix element)
involving that state. In a sense, DMRG is asked to work harder than
necessary: it is used to calculate a single state with
``general-purpose accuracy'', whereas the accurate calculation of a
particular expectation value involving that state would have been
sufficient.

The main motivation for the present work is to attempt to reduce this
calculational overhead by employing a representation of the spectral
function that avoids the need for calculating a single state with such
high accuracy and instead allows numerical resources to be focussed
directly on the calculation of the relevant expectation values.  This
can be achieved by representing the spectral function via a
\emph{Chebychev expansion},\cite{Wheeler1974,SilverRoeder1994,Weisse2006} whose
coefficients, the so-called \cheby moments, can be calculated
recursively using MPS tools. Below, we briefly summarize the structure
and main features of such an expansion, thereby providing both an
introduction and an overview of the material developed in detail in
the main part of this paper.

The Chebychev polynomials $T_n(x)$ form an orthonormal set of
polynomials on the interval $x \in [-1, 1]$.  They are very well
studied mathematically,\cite{AbramowitzStegun1970,Boyd1989,Rivlin1990}
and are widely used for function expansions since they have very
favorable convergence properties. As will be described in detail
below, the spectral function can be represented approximately by a
so-called Chebychev expansion, which becomes exact for $N\to \infty$,
of the following form:
\begin{equation}
  \label{eq:spectral-exp-A-intro}
  \mathcal{A}^{\mathcal{BC}}_N(\omega)
  =  \frac{2W'/\Wast}{\pi \sqrt{1-\wshift^{2}}}
  \left[g_0 \mu_{0} + 2\sum_{n=1}^{N-1}  g_n \mu_{n}T_{n}(\wshift)\right].
\end{equation}
Here the \emph{\cheby moments} $ \mu_{n} = \bra{0}\Bop\ket{t_{n}}
$ are obtained from the \emph{Chebychev vectors} $ \ket{t_{n}} =
T_{n}(\hat \Hshift)\Cop\ket{0}$, and the $g_n$ are known \emph{damping
  factors} that influence broadening effects. The primes indicate
that the Hamiltonian $\hat H$ and frequency $\omega$ were expressed in
terms of rescaled and shifted versions, $\hat H'$ and $\omega'$, in
such a manner that an interval $\omega \in [0, \Wast]$ that contains
the entire spectral weight is mapped onto a \emph{rescaled band}
$\omega' \in [-W',W']$ of halfwidth $W' < 1$.

\begin{figure}[t]
  \centering
  \includegraphics[width=\linewidth]{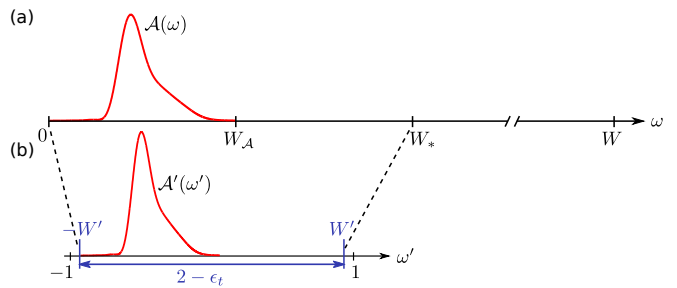}
  \caption{(a) Sketch of a spectral function whose spectral width
      $\WA$ is much smaller than the many-body bandwidth $W$. Before
      making a Chebychev expansion, we rescale the interval $\omega
      \in [0, \Wast]$, with effective bandwidth $\Wast = 2 \WA$, onto
      the interval $\omega' \in [-W', W']$, shown in (b), with
      rescaled half-bandwidth $W' = 1 - \frac{1}{2} \epsilont$ and a
      safety factor $\epsilont \simeq 0.025$.}
  \label{fig:rescaling-sketch}
\end{figure}

This representation has several useful features: \\
(i) It resolves the interval $\omega \in [0, \Wast]$ with a \emph{uniform}
resolution of $\mathcal{O}(\Wast/N)$.\\
(ii) The range of frequencies over which the spectral function has
nonzero weight, say $\WA$ (to be called its \emph{spectral width}) is
often significantly smaller than the many-body bandwidth of the
Hamiltonian, say $W$, as depicted in Fig~\ref{fig:rescaling-sketch}.
By choosing the effective bandwidth $\Wast$ to be of order $\WA$
instead of $W$, huge gains in resolution are possible.
\\
(iii) A well-controlled broadening scheme, encoded in the damping
factors $g_n$, is available that allows finite-size effects to be
either resolved or smeared out, as desired.
\\
(iv) The \cheby vectors $|t_n \rangle $ are calculated using a
(numerically stable) \emph{recursion} scheme, which exploits Chebychev
recurrence relations to calculate $|t_n\rangle$ from $H'|t_{n-1}
\rangle$ and $|t_{n-2} \rangle$ (see \Eq{eq:dmrg-recipe-3}). Thus, the
expectation values from which the spectral function is constructed are
built up in a series of recursive steps (see \Eq{eq:sum-mu} below)
instead of being calculated at the end in one final step.\\
(v) The bond entropy of successive \cheby vectors $|t_n \rangle$
is found empirically to \emph{remain bounded} with increasing
recursion number $n$, thus the complexity of these vectors remains
managable up to arbitrarily large $n$.\\
(vi) Finally, and from the perspective of numerical costs, most
importantly: \CheMPS \emph{efficiently copes} with the \emph{growth in
  bond entropy} with increasing iteration number that usually limits
DMRG approaches. It does so by distributing this entropy over all
$|t_n \rangle$, thereby packaging it into managable units (see (v)).
In particular, when constructing and using the states $ |t_n \rangle$,
one never needs to know more than three at a time (and after use may
delete them from memory).  Hence, it is not necessary to combine all
information contained in all $|t_n \rangle$ into a single MPS.  

Let us constrast this with the CV or tDMRG approaches: imagine
expanding the correction vector or time-evolved state in terms of the
\cheby vectors $|t_n \rangle$, i.e.\ expressing them as linear
combinations of the form
\begin{eqnarray}\label{eq:lin-comb}
  |\mathcal{C} \rangle_\omega \simeq \sum_{n=0}^{N-1} C^n_\omega
  |t_n \rangle \; , \quad 
  |\tilde {\mathcal{C}}\rangle_t \simeq  \sum_{n=0}^{N-1} 
  \tilde C^n_t 
  |t_n \rangle \; , 
\end{eqnarray}
respectively. (The coefficients $C^n_\omega$ and $\tilde C^n_t $
are related by Fourier transformation.)
Now, the CV or tDMRG approaches in effect attempt to
accurately represent the \emph{entire linear combination} using a
single MPS. This endevour is numerically very costly, since the
entanglement entropy of this linear combination grows rapidly with
$N$. The Chebychev approach avoids this problem by taking expectation
values \emph{before} performing the sum on $n$:
\begin{eqnarray}\label{eq:sum-mu}
  \langle 0  | \Bop 
  |\Cop\rangle_\omega \simeq \sum_{n=0}^{N-1} C^n_\omega
  \mu_n  \; , \quad 
  \langle 0 | \Bop  |\tilde {\mathcal{C}}
  \rangle_t \simeq  \sum_{n=0}^{N-1} 
  \tilde C^n_t \mu_n \; . 
\end{eqnarray}
Thus, the Chebychev expansion very conveniently organizes the calculation
into many separate, and hence numerically less costly, packages or subunits. 

Our paper is organized as follows. We introduce the \cheby expansion
for spectral functions in \fref{sec:chebyshev-expansion} and discuss
its implementation using MPS including a new algorithm for performing
a projection in energy in \fref{sec:mps-eval-coeffs}. In
\fref{sec:heisenberg} we present \CheMPS results for the structure
factor of a spin-$\frac{1}{2}$ Heisenberg chains, perform a detailed
analysis of finite-size effects (see
\fref{fig:finite-size-analysis}), and compare our results to CV,
Bethe Ansatz and tDMRG (see Figs.~\ref{fig:spin-cv-compare},
\ref{fig:spin-L-compare} and \ref{fig:time-skt}, respectively).  In
Sec.~\ref{sec:error-analysis} we perform an extensive error analysis
of the \CheMPS approach using the quadratic resonant level model,
and discuss some salient features of density matrix eigenspectra
in Sec.~\ref{sec:density-matrix}.  Section~\ref{sec:conclusion}
summarizes our main conclusions, and Sec.~\ref{sec:outlook} presents
a brief outlook towards possible future applications, involving time
dependence or finite-temperature correlators. An Appendix gives a
detailed account of \CheMPS results for the resonant level model
used for the error analysis of Sec.~\ref{sec:error-analysis}.

\section{\cheby expansion of $\mathcal{A}^{\mathcal{BC}} (\omega)$}
\label{sec:chebyshev-expansion}

\subsection{\cheby basics}
\label{sec:cheby-basics}

Let us start by briefly summarizing those properties of \chebyps
that will be needed below.  We follow the notation of
Ref.~\onlinecite{Weisse2006}, which gives an excellent general
discussion of \cheby expansion techniques (though without mentioning
possible DMRG/MPS applications).

\begin{figure}[tb]
  \centering
  \includegraphics[width=\linewidth]{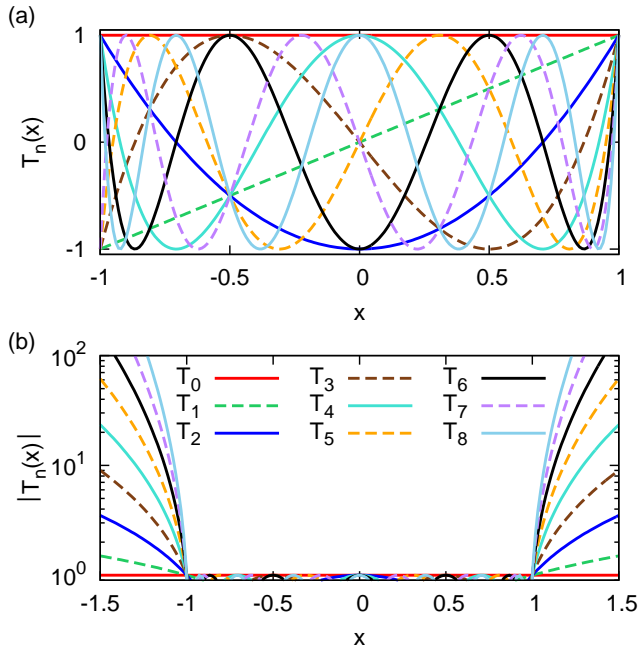}
  \caption[\chebyps of the first kind $T_{n}(x)$ for $n$ up to
  $8$.]{(Color online) \chebyps of the first kind, $T_{n}(x)$, for $n$
    up to $8$. (a) All zeros and extrema of every $T_{n}(x)$ are
    located within the interval $I = [-1,1]$, and all extremal values
    equal $1$ or $-1$. (b) \chebyps $|T_{n}(x)|$ for $x \in
    [-1.5,1.5]$. The $|T_{n>0}(x)|$ grow rapidly when $\abs{x}$
    increases beyond 1.}
  \label{fig:app-cheby-poly}
\end{figure}

\chebyps of the first kind, $T_{n}(x)$, henceforth simply called
\chebyps, are defined by the recurrence relations
\begin{equation}
  \label{eq:cheby-recursion}
  \begin{split}
    T_{n+1}(x) &= 2 x T_{n}(x) - T_{n-1}(x),\\
    T_{0}(x) &= 1,\qquad T_{1}(x) = x \; .
  \end{split}
\end{equation}
They also satisfy the useful relation (for $n\ge n'$)
\begin{equation}
  \label{eq:cheby-doublerecursion}
  T_{n+n'}(x)  =   2 T_{n}(x) T_{n'}(x) - T_{n-n'}(x)\; .
\end{equation}
Two useful explicit representations are: 
\begin{equation}
  \label{eq:cheby-poly-explicit}
  T_{n}(x) = \cos\left[n \arccos(x)\right] =  
  \cosh\left[n \, \textrm{arccosh}(x)\right] .
\end{equation}
On the interval $I = [-1,1]$ the \chebyps constitute an orthogonal
system of polynomials (over a weight function $(\pi \sqrt{1-x^2})^{-1}$), in
terms of which any piecewise smooth and continuous function $\left.
  f(x)\right|_{x\in I}$ can be expanded.  In fact, the $T_n(x)$ are
optimally suited for this purpose, since they have the unique property
(setting them apart from other systems of orthogonal polynomials) that
on $I$ their values are confined to 
$|T_n(x)| \le 1$, with all extremal values equal to $ 1$ or $-1$.
This is evident from the first equality in
Eq.~(\ref{eq:cheby-poly-explicit}); the second equality implies that
for $x \notin I$, $|T_n(x)|$ grows rapidly with increasing
$|x|$. These properties are illustrated in
Fig.~\ref{fig:app-cheby-poly}.

There are several ways of constructing \cheby approximations for
$\left. f(x) \right|_{x\in I}$ (see Weisse et~al.,\cite{Weisse2006},
Section II.A).  The \cheby expansion that is practical for present
purposes has the form
\begin{align}
  \label{eq:cheby-expansion-infinite}
  f(x) &= \frac{1}{\pi\sqrt{1-x^{2}}} 
  \left[\mu_{0} + 2\sum_{n=1}^{\infty} \mu_{n} T_{n}(x) \right] \; , 
\end{align}
where the \emph{Chebychev moments} $\mu_n$ are 
given by 
\begin{align}
  \label{eq:cheby-expansion-mu}
  \mu_{n} &= \int_{-1}^{1} d x f(x) T_{n}(x) \; .
\end{align}
An approximate representation of order $N$ is obtained for $f(x)$ if
only the first $N$ terms (i.e.\ $n \le N-1$) are retained. However,
such a truncation in general introduces artificial oscillations, of
period $\simeq 1/N$, called \emph{Gibbs oscillations}.  These
can be smoothened by employing certain broadening kernels, 
which in effect rearrange the infinite
series (\ref{eq:cheby-expansion-infinite}) before truncation. 
This leads to a \emph{reconstructed} expansion of the form
\begin{align}
  \label{eq:cheby-expansion-f}
  f_{N}(x) &= \frac{1}{\pi\sqrt{1-x^{2}}} \left[g_0 \mu_{0} +
    2\sum_{n=1}^{N-1} g_n \mu_{n} T_{n}(x)\right] \; ,
%\\
%  \label{eq:cheby-expansion-mu}
%  \tilde{\mu}_{n} &= g_{n}\mu_{n} = g_{n}\int_{-1}^{1}f(x) T_{n}(x)
%  dx,\qquad \tilde{\mu}_{n} = g_{n}\mu_{n}\,,
\end{align}
which (for properly chosen kernels) converges \emph{uniformly}:
\begin{align}
  \label{eq:uniform-convergence}
  \max_{-1 < x < 1} |f(x) - f_N(x)| \stackrel{N \to \infty}{\longrightarrow}
   0 \; .
\end{align}
The reconstructed series (\ref{eq:cheby-expansion-f}) contains
the same \cheby moments $\mu_n$ as \Eq{eq:cheby-expansion-mu}, but they are
multiplied by \emph{damping factors} $g_n$, real numbers whose form is characteristic of the
chosen kernel. Several choices have been proposed, which damp out
Gibbs oscillations in somewhat different ways
(see Ref.~\onlinecite{Weisse2006} for
details). We will mostly employ \emph{Jackson damping}, given by
\begin{equation}
  \label{eq:app-cheby-12}
  g_{n}^{J} = \frac{(N-n+1)\cos\frac{\pi n}{N+1} + 
    \sin\frac{\pi n}{N+1}\cot\frac{\pi}{N+1}}{N+1} .
\end{equation}
This is usually the best choice, since it guarantees an integrated
error of $\mathcal{O}(\frac{1}{N})$ for $f_{N}(x)$.  When used to
approximate a $\delta$-function $\delta(x-\bar x)$ sitting at
$\bar x \in I$, Jackson damping yields a nearly Gaussian peak of
width $\sqrt{1- \bar x^2} \, \pi/N$. On one occasion we will also
employ \emph{Lorentz damping},
\begin{equation}
  \label{eq:app-cheby-13}
  g_{n,\lambda}^{L} = \frac{\sinh\left[\lambda \left(1-\frac{n}{N} \right)\right]}{\sinh\lambda},
\end{equation}
where $\lambda$ is a real parameter. 
Lorentz damping preserves analytical properties (causality) of Green's
function and broadens a $\delta$-function $\delta(x-\bar x)$ into a peak
whose shape, for the choice $\lambda=4$ used here (following
Ref.~\onlinecite{Weisse2006}), is nearly Lorentzian, of width
$\sqrt{1-\bar x^2} \, \lambda/N$.

To summarize: the order-$N$ Chebychev-reconstruction $f_N(x)$ with
Jackson or Lorentzian damping with $\lambda = 4$ yields a result that
is very close to the broadened function
\begin{eqnarray}
  \label{eq:effective-broadening}
  f^{X}_N (x) = \int_{-1}^1 d \bar x K_{\eta'_{N, \bar x}}^X (x - \bar x) f(\bar x) \; , 
\end{eqnarray}
($X = J,L$) with broadening kernels and widths given by
\begin{subequations}
  \label{eq:cheby-kernels}
  \begin{eqnarray}
    \label{eq:kernels-J}
    K^J_{\eta'} (x)  & = & \frac{e^{-{x^2}/(2 (\eta'^2) }}{\sqrt{2 \pi} \eta'} \; , \qquad
    \eta'_{N,\bar x}  = \sqrt{1- \bar x^2} \, \frac{\pi}{N} \; , \qph \\
    \label{eq:kernels-L}
    K^L_{\eta'} (x)  & = & \frac{\eta' / \pi}{x^2 + {\eta'}^2} \; , \qquad
    \eta'_{N,\bar x} = \sqrt{1- \bar x^2} \, \frac{4}{N} \; ,
  \end{eqnarray}
\end{subequations}
respectively. Thus, $f_N(x)$ resolves the shape of $f(x)$ with a
resolution of $\mathcal{O}(1/N)$.  

\begin{figure}[tb]
  \centering
  \includegraphics[width=\linewidth]{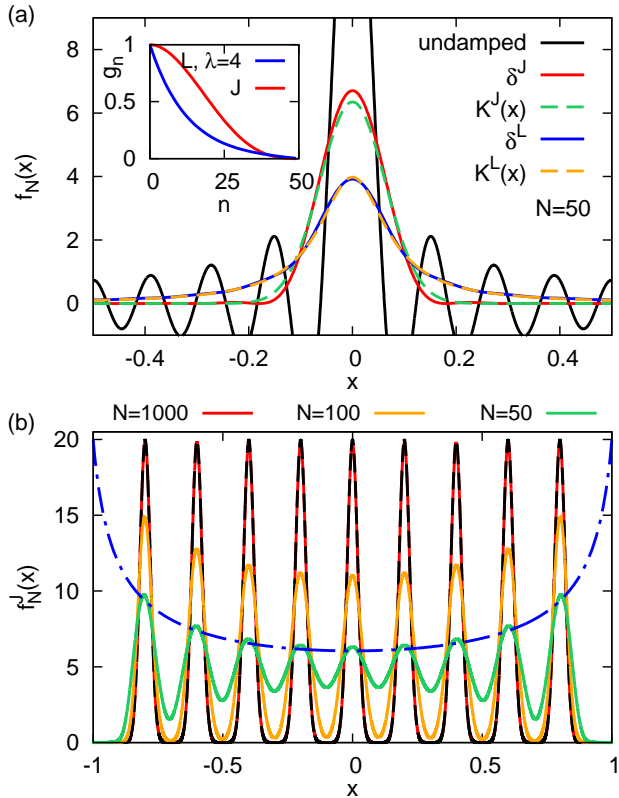}
  \caption[\cheby expansion of $\delta(x)$]{Three \cheby
    reconstructions of $\delta(x)$, with $N=50$: the undamped case
    $(g_n = 1)$ yields Gibbs oscillations (central peak has height
    $\delta_{50}(0) = 16.23$); Jackson damping $(\delta^J)$ mimicks a
    Gaussian peak $K^J(x)$ of width $\pi/N$; Lorentz damping
    ($\delta^L$) for $\lambda = 4$ mimicks a Lorentzian peak $K^L(x)$
    of width $\lambda/N$. Inset: Jackson and Lorentz damping factors,
    $g_n^J$ and $g_{n,\lambda=4}^L$, respectively, plotted for
    $N=50$. Both decrease monotonically from 1 to 0, but in somewhat
    different ways. (b) Jackson-damped reconstruction of a comb
    of normalized Gaussians (dashed line), all of width $\bar \eta' =
    0.02$, for three values of $N$ (solid lines). The $
    x$-dependence of the  peak heights is given by $[2 \pi(
    \bar  \eta^{\prime 2}+ \eta^{\prime 2}_{N,x})]^{-1/2}$ (dash-dotted line), see
    \Eq{eq:adding-Gaussian-widths}.}
  \label{fig:kernel-delta}
\end{figure}

For purposes of illustration, \fref{fig:kernel-delta}(a) shows three
\cheby reconstructions of a $\delta$-function at $\bar x = 0$: without
damping, giving Gibbs oscillations; with Jackson damping, yielding a
near-Gaussian peak; and with Lorentz damping, yielding a
near-Lorentzian peak. Figure~\ref{fig:kernel-delta}(b) shows a
Jackson-damped \cheby reconstruction of a comb of Gaussian peaks
$\sum_\alpha K^J_{\bar \eta'_\alpha}(x - \bar x_\alpha)$, whose
widths $\bar \eta'_\alpha$ are all equal. It illustrates how
increasing $N$ reduces the amount of broadening until the original
peak form is recovered for sufficiently large $N$. It also shows
that the broadened peak widths depend on the peak positions, 
reflecting the fact that convolving a Gaussian of width $\bar \eta'_\alpha$
with a near-Gaussian of width $ \eta'_{N, \bar x_\alpha}$
(\Eq{eq:effective-broadening}) produces a near-Gaussian of width
\begin{eqnarray}
  \label{eq:adding-Gaussian-widths}
  \eta'_\alpha \simeq \sqrt{\bar \eta^{\prime 2}_\alpha + 
    \eta^{\prime 2}_{N, \bar x_\alpha}} \; .
\end{eqnarray}

To evaluate the $T_{n}(x)$ that occur in \Eq{eq:cheby-expansion-f}, we
use the first equality of Eq.~\eqref{eq:cheby-poly-explicit}.
Although numerically more efficient methods exist for this
purpose,\cite{Weisse2006} their use becomes advisable only for
expansion orders much larger than the $N \lesssim {\cal O}(10^{3})$
that we will need in this work.

\subsection{Rescaling of $\omega$ and $\hat H$}
\label{sec:rescaling-H}

To construct a \cheby expansion 
of the spectral function $ \mathcal{A}^{\mathcal{BC}}(\omega)$ of
\Eq{eq:spectral-frequency}, we need to rescale and shift
\cite{Weisse2006} the Hamiltonian  $\hat H \mapsto \hat H'$ and the
frequency $\omega \mapsto \omega'$ in such a way that the spectral
range of $\mathcal{A}(\omega)$, i.e.\ the interval $[0, \WA]$ within
which it has nonzero weight, is mapped \emph{into} the interval
$[-1,1]$. Rescaled, dimensionless energies and frequencies will
always carry primes.  As safeguards against ``leakage'' beyond
$[-1,1]$ due to numerical inaccuracies, we choose the linear map (see
\figref{fig:rescaling-sketch})
\begin{eqnarray}
  \label{eq:mapping-intervals}
  \omega \in [0, \Wast] \mapsto \omega' \in [-
  W', W'] \; , \quad 
  W' = 1 - {\textstyle \frac{1}{2}} \epsilont \; , \qph
\end{eqnarray}
which entails two precautionary measures: first, the
$\omega$-interval is taken to be larger than the requisite $[0, \WA]$
by choosing the \emph{effective bandwidth} $\Wast$ to be larger than
the spectral width $\WA$; second, the $\omega'$-interval is taken to
be slightly smaller than the requisite $[-1,1]$ by choosing the
\emph{rescaled half-bandwidth} $W'$ to be smaller than 1, with a
safety factor\cite{Weisse2006} of $\epsilont \simeq 0.025$.  To be
explicit, we define
\begin{subequations}
  \label{eq:mapping-omega-H}
  \begin{eqnarray}
    \label{eq:mapping-omega}
    \omega' & = & \frac{\omega}{a} - W' \; , 
    \quad a = \frac{\Wast}{2 W'} \; , \\
    \label{eq:cheby-h-rescaled} 
    \hat H' & = & \frac{\hat H-E_0}{a} - W' \; ,
  \end{eqnarray}
\end{subequations}
where $\hat H'$ has ground state energy $E_0' = - W'$. Then
we express the spectral function (\ref{eq:spectral-frequency}) as
\begin{equation}
  \label{eq:exp-deriv-3}
  \mathcal{A}^{\mathcal{BC}}(\omega) = \frac{1}{a} 
  \bra{0}\Bop
  \, \delta (\wshift - \hat H') \, \Cop\ket{0} \; , 
\end{equation}
(with $\omega'= \omega'(\omega)$ and $\hat H'$ given by
\Eqs{eq:mapping-omega-H}), which by construction has no weight for
$\omega' \notin [- W', W']$.

One possible choice for $\Wast$ is to equate it to the width of the
many-body spectrum of $H$, given by $W = E_{\rm max} - E_{0}$.  When
using DMRG, $E_{0}$ is usually already known from calculating the
ground state $\ket{0}$ of $H$, and $E_{\rm max}$ can be found, \eg, by
calculating \footnote{In principle, the Lanczos algorithm also
  provides the maximal eigenvalue. However, in DMRG the Lanczos gets
  restarted at every site with the currently known ground state and
  thus the return maximal energy will no longer approach $E_{\rm max}$
  as the ground state converges.} the ground state of $-H$ (reduced
DMRG accuracy relative to usual ground state calculations is
sufficient, since only $E_{\rm max}$ is of interest here.)

A disadvantage of the choice $\Wast = W$ is that the many-body
bandwidth $W$ typically is large (it scales with system size), whereas
optimal spectral resolution requires $\Wast$ to be as
small as possible: since an $N$-th order \cheby expansion yields a
resolution of $\mathcal{O}(1/N)$ on the interval $[-1,1]$, its
resolution on the original interval $[0, \Wast]$ will be
$\mathcal{O}(\Wast/N)$, which evidently becomes better the smaller
$\Wast$.  If $\Bop$ and $\Cop$ are single-particle
operators, the spectral width $\WA$ of
$\mathcal{A}^{\mathcal{BC}}(\omega)$ is independent of system size and
hence much smaller than the many-body bandwidth $W$.  In this case, it
is advisable to choose $\Wast$ to be of similar order (though still
larger) than $\WA$. We will choose $\Wast = 2\WA$, which 
is typically $\ll W$, as illustrated in \fref{fig:rescaling-sketch}.

\subsection{Chebychev expansion in frequency domain}
\label{sec:expansion-A}

To expand the $\delta$-function in \Eq{eq:exp-deriv-3} in \chebyps,
we use $\hat f(x) = \delta (x - \hat H')$ with $x = \omega'$ in
\Eq{eq:cheby-expansion-mu}, and obtain from
Eq.~\eqref{eq:cheby-expansion-f} a reconstructed \cheby operator expansion of
the form:
\begin{equation}
  \label{eq:spectral-exp-A}
  \delta_N(\wshift - \hat H') = 
  \frac{1}{\pi \sqrt{1-\wshift^{2}}}
  \left[g_0  + 2\sum_{n=1}^{N-1} g_n T_n (\hat H') T_{n}(\wshift)\right] .
\end{equation}
Inserting this into \Eq{eq:exp-deriv-3} for $\mathcal{A}^{\mathcal{BC}}(\omega)$
yields the \cheby expansion (\ref{eq:spectral-exp-A-intro}), with
\cheby moments given by
\begin{eqnarray}
  \label{eq:spectral-exp-mu-res}
  \mu_{n} & = & \bra{0} \Bop T_{n}( \hat \Hshift)\Cop\ket{0} \; .
\end{eqnarray}
Thus $\mu_{n}$ is a ground state expectation value of an $n$-th order
polynomial in $\hat \Hshift$, whose construction might \emph{a priori}
appear to become increasingly daunting as $n$ increases. Fortunately,
this challenge can be dealt with \emph{recursively}, by expressing the
moments as
\begin{equation}
  \label{eq:dmrg-recipe-1}
  \mu_{n} = \bra{0}\Bop\ket{t_{n}} \; , \qquad
  \ket{t_{n}} = T_{n}(\hat \Hshift)\Cop\ket{0},
\end{equation}
and calculating the \emph{\cheby vectors} $\ket{t_{n}}$
by exploiting the \cheby recurrence relations
(\ref{eq:cheby-recursion}).  The details of this recursive scheme will
be discussed in Section~\ref{sec:mps-eval-coeffs}.

\subsection{Chebychev expansion in time domain}
\label{sec:expansion-G-BCt}

The \cheby expansion can also be employed for studying time evolution
in general, and the correlator $G^{\mathcal{BC}}(t)$ in particular.
To this end, we express the time-evolution operator as
\begin{eqnarray}
  \label{eq:U-cheby}
   \hat  U(t) &= & e^{-i\hat Ht} = 
   \int_{-1}^1 d \wshift e^{-i\left[a(\wshift +W') + E_{0}\right]t}
   \delta (\wshift - \hat H') \; , \qqph 
\end{eqnarray}
and insert \Eq{eq:spectral-exp-A} (without damping, $g_n = 1$) into
the latter. This yields\cite{Tal-Ezer1984,Leforestier1991}
\begin{subequations}
\label{eq:U-cheby-all}
\begin{eqnarray}
  \label{eq:U-time-cheby}
  \hat U_N (t)    &= &  e^{-i(E_{0} + aW')t} \left[c_{0}(t) + 
    2 \sum_{n=1}^{N-1} T_{n}(\hat \Hshift)  c_{n}(t) \right] \!\! , \qqph \\
  \label{eq:c-time-cheby}
  c_{n}(t) &= &\int_{-1}^{1} \frac{e^{-iat \wshift} T_{n}(\wshift)}
  {\pi\sqrt{1-\omega^{\prime 2}}} d \wshift 
  = (-i)^{n} J_{n}(at) \, .
\end{eqnarray}
\end{subequations}
Here $J_{n}(at)$ is the Bessel function of the first kind of
order $n$. It decays very rapidly with $n$ once $n > at$.  Hence, an
expansion of given order $N$ gives an essentially exact
representation of $\hat U(t)$ for times up to $t_{\rm max} \lesssim
\frac{N}{a}$, while $c_{N-1}(t)$ provides an estimate of the error.

Inserting \Eqs{eq:U-cheby-all}  into \Eqs{eq:tDMRG}  for 
$G^{\mathcal{BC}}(t)$ we find
\begin{equation}
  \label{eq:cheby-Gt}
  G_N^{\mathcal{BC}}(t) = e^{-i aW't} 
  \left[\mu_0 J_{0}(at) + 2 \sum_{n=1}^{N-1} (-i)^n \mu_n \, J_n (at)  \right] ,   
\end{equation}
where the \cheby moments $\mu_n$ are again given by
\Eq{eq:dmrg-recipe-1}.  Thus the \cheby expansions of
$G^{\mathcal{BC}}(t)$ and $ \mathcal{A}^{\mathcal{BC}} (\omega)$ are
governed by the \emph{same} set of moments $\mu_n$, as is to be
expected for functions linked by Fourier transformation.

\section{MPS evaluation of the \cheby moments $\mu_{n}$}
\label{sec:mps-eval-coeffs}

We now present a recursive scheme for calculating the \cheby moments
$\mu_{n}$.  The manipulations described below were implemented using
MPS-based methods,\cite{dukelsky1998,Vidal2004,verstraete:dmrg-pbc,Verstraete2004b,McCulloch2007,Schollwoeck2010}
which are very convenient for constructing the states of interest,
while matrix-product operators\cite{McCulloch2007} (MPOs) simplify the
implementation of the shift- and rescaling transformation
\fref{eq:cheby-h-rescaled} of the Hamiltonian.

\subsection{Recurrence fitting}
\label{sec:cheby-iteration-step}

To initialize the \cheby expansion, we calculate ground state
$\ket{0}$ and ground state energy $E_0$ of $\hat H$,  make a specific
choice for $\Wast$ and $W'$, and construct $\hat \Hshift$ according to
Eq.~\eqref{eq:cheby-h-rescaled}.  Then comes the main task, namely the
recursive calculation of the moments $\mu_{n}$.  This is done starting from
\begin{equation}
  \label{eq:dmrg-recipe-2}
  \ket{t_{0}} = \Cop\ket{0},\qquad \ket{t_{1}} = \hat \Hshift\ket{t_{0}} \; , 
\end{equation}
and using the recurrence relation (obtained from
Eq.~\eqref{eq:cheby-recursion})
\begin{equation}
  \label{eq:dmrg-recipe-3}
  \ket{t_{n}} = 2 \hat \Hshift \ket{t_{n-1}} - \ket{t_{n-2}} \; .
\end{equation}
\Eq{eq:dmrg-recipe-3} can be implemented using the so-called compression
or fitting procedure\cite{Verstraete2004} (see \cite{Schollwoeck2010},
Sec.~4.5.2 for details). It finds an MPS representation for $|t_n \rangle$, 
at minimal loss of information for given MPS dimension $m$, by
variationally minimizing the \emph{fitting error}
\begin{equation}
  \label{eq:fitting-norm}
  \Delta_{\rm fit} =
  \left\|\ket{t_{n}} - \left(2\hat \Hshift\ket{t_{n-1}} - \ket{t_{n-2}}\right)\right\|^{2} \; . 
\end{equation}
We will call this procedure \emph{recurrence fitting}. In practice,
the variational minimization proceeds via a sequence of fitting sweeps
back and forth along the chain. These are continued until the state
being optimized becomes stationary, in the sense that the 
overlap 
\begin{equation}
  \label{eq:fitting-convergence}
  \Delta_{\rm c} = \left| 1 - \frac{\bracket{t_n}{t'_n}}{\norm{\ket{t_n}}\norm{\ket{t'_n}}}\right| 
\end{equation}
between the states $|t_n\rangle$ and $|t'_n \rangle$ before and
after one fitting sweep, drops below a specified \emph{fitting
  convergence threshold} (typically in the range $10^{-6}$ to
$10^{-8}$). The maximum expansion order for which $|t_n \rangle$
is obtained using recurrence fitting will be denoted by $\Nmax$.

The MPS dimension $m$ needed to achieve accurate recurrence fitting
turns out to be surprisingly small (see \fref{sec:error-analysis} for
a detailed analysis). For example, $m = 32$ sufficed for the
antiferromagnetic Heisenberg chain of length $L=100$ discussed in
Section~\ref{sec:heisenberg}. The reason for this remarkable and
eminently useful feature lies in the fact that the Chebychev
recurrence relations (\ref{eq:dmrg-recipe-3}) contain only two terms
on the right-hand side, whose addition requires only modest
computational effort.  In contrast, CV or tDMRG typically require much
larger $m$, since they attempt to represent the sum of many states,
see \fref{eq:lin-comb}, in terms of a single MPS.

For the special but common case that
$\Bop = \Cop^{\dagger}$, Eq.~\eqref{eq:cheby-doublerecursion} 
yields a relation between different moments, 
\begin{equation}
  \label{eq:moments-consistent}
  \mu_{n+n'} = 2\bracket{t_{n}}{t_{n'}} - \mu_{n-n'} \; . 
\end{equation}
This can be used to effectively double the order of the expansion
to $2 \Nmax$ without calculating any additional \cheby vectors, by setting
$n' = n-1$ or $n$:
\begin{equation}
  \label{eq:dmrg-recipe-4}
  \begin{split}
   \tilde \mu_{2n-1} &= 2\bracket{t_{n}}{t_{n-1}} - \mu_{1},\\
   \tilde \mu_{2n} &= 2\bracket{t_{n}}{t_{n}} - \mu_{0}.
  \end{split}
\end{equation}
We use tildes to distinguish $\tilde \mu_n$-moments calculated in this
manner from the $\mu_n$-moments obtained via
\Eq{eq:dmrg-recipe-1}. Although they should nominally be identical, in
numerical practice $\tilde \mu_n$-moments are less accurate (by up to
a factor of 5 in \fref{fig:rlm-coeffs-illustrate}(c) below), since
they depend on two \cheby vectors, whereas $\mu_n$-moments depend on
only one.  Our \cheby reconstructions thus generally employ the
$\mu_n$-moments, and unless stated otherwise, $\tilde \mu_n$-moments
are used only for results requiring $\Nmax \le n < 2\Nmax$.

\subsection{Energy truncation}
\label{sec:dmrg-energy-trunc}

\begin{table*}
  \centering
  \begin{ruledtabular}
    \begin{tabular}{cccc}
      parameter & recommended value & description & task \\
      \hline
      $\Wast$ & $ 2 W_{\mathcal{A}} $ (or $W$) &  effective bandwidth with (or without) energy truncation & \multirow{2}{*}{rescaling of $H$}\\
      $\epsilont$ & $0.025$ & safety offset in rescaled half-bandwidth: 
      $W' = 1 - \frac{1}{2} \epsilont$ & \\
      \hline
      $m$ & & MPS dimension & \multirow{2}{*}{recurrence fitting}\\
      $\Delta_{\rm c}$ & $10^{-6}\ldots10^{-8}$ & fitting convergence threshold & \\
      \hline
      $\dtrunc$ & $30$ & Krylov subspace dimension & \multirow{3}{*}{energy truncation}\\
      $\ntrunc$ & $10$ & number of sweeps & \\
      $\etrunc$ & $1.0$ & energy truncation threshold
      (in rescaled units)  & \\
      \hline
      $N$ & depends on system size & order of expansion, broadening & \multirow{2}{*}{spectral reconstruction}\\
      $g_{n}$ & $g_{n}^{J}$ & choice of damping factors & \\
    \end{tabular}
  \end{ruledtabular}
  \caption[List of parameters that influence the quality of the 
  DMRG expansion]{List of \CheMPS parameters that control 
    various algorithmic tasks, influencing their numerical
    costs and the quality of results.
    Since  the tasks ``recurrence fitting'' and ``energy 
    truncation'' are carried out at every recursion step, the importance
    of the corresponding parameters is self-evident. However, $\Wast$ 
    and $\epsilont$, which determine the rescaled \ham $\hat H'$, turn out 
    to have a high impact on the results, too,
    as the quality of the energy truncation
    sweeps strongly depends on $\hat H'$. For $\etrunc=1$,  
    the choice of taking $\Wast$ to be 
    twice the spectral bandwidth $W_{\mathcal A}$ 
    (or equal to the many-body bandwidth $W$)
    was found to work well with (or without) energy truncation, respectively.
    $N$ and $g_{n}$ do not affect the calculated moments of the
    expansion, but control the broadening of the reconstructed spectral 
    function.}
  \label{tab:expansion-params}
\end{table*}

We have argued above that in order to optimize spectral resolution, it
may be desirable to choose the effective bandwidth $\Wast$ to be
smaller than the full many-body bandwidth $W$.  If this is done,
however, it is essential to include an additional \emph{energy
  truncation} step into the recursion procedure, to ensure that each
$\ket{t_{n}}$ remains free from ``high-energy'' components, i.e.\
$\hat H'$-eigenstates with eigenenergies $E_k' > 1$, which fall
outside the range $[-1,1]$ that is admissable for arguments of \cheby
polynomials.  If $\Wast < W$, numerical noise causes the state
$\ket{t_{n}}$ to contain such high-energy contributions in spite of
the precautionary measures described after \Eq{eq:mapping-intervals},
because the application of $\hat H'$ to $|t_{n-1} \rangle$ in
\Eq{eq:dmrg-recipe-3} entails a DMRG truncation step, which is not
performed in the eigenbasis of $\hat H'$.  If such high-energy
components were fed into subsequent recursion steps, the norms $\norm{
  t_n \rangle}$ of successive \cheby vectors would diverge rapidly (as
would the resulting moments $\mu_{n}$), because this effectively
amounts to evaluating \chebyps $T_{n}(x)$ for $\abs{x} > 1$, where
$|T_{n}| \gg 1$ (see \figref{fig:app-cheby-poly}(b)).

As a consequence, after obtaining a new state $\ket{t_{n}}$ from
Eq.~\eqref{eq:dmrg-recipe-3}, we take the precautionary measure of
projecting out any high-energy components that it might contain,
before proceeding to the next $|t_{n+1} \rangle$.  This can be done by
performing several \emph{energy truncation sweeps}. During an energy
truncation sweep, we focus on one site at a time, perform an energy
truncation in a local Krylov basis constructed for that site, and then
move on to the next site. Shifting the current site is accomplished by
standard MPS means, without any truncation, as a DMRG truncation would
counteract the energy truncation. (As a consequence, an energy
truncation in terms of two-site sweeps has not been implemented.)

The truncation must take place in the energy eigenbasis of the \ham
$\hat H'$. Of course, its complete eigenbasis is not accessible,
thus we build a Krylov subspace of dimension $\dtrunc$ within the
effective Hilbert space at every site. Alternatively, energy
truncation can also be performed in the bond representation
$\ket{\psi} = B_{lr}\ket{l_{k}}\ket{r_{k}}$. In this Krylov
subspace, the effective \ham $\hat H'_{K}$ of dimension
$\dtrunc$ can be fully diagonalized and so we can construct a
projection operator to project out all eigenstates with energy
bigger than some \emph{energy truncation threshold} $\etrunc$. The
choice of this threshold depends on the choice of $\Wast$. We have
found the combination $\Wast = 2 W_A$ and $\etrunc = 1.0$ to work
well (but other choices, involving, e.g.\ smaller $\Wast$ and larger
$\etrunc$ would be possible, too.)

In the following, we describe the procedure just outlined in more
detail for a single site, using standard MPS nomenclature. Let
the effective local Hilbert space for this site be spanned by the
left, local and right basis vectors $\ket{l}$, $\ket{\sigma}$ and
$\ket{r}$, and expand the \cheby vector $\ket{\psi} = \ket{t_{n}}$
in this basis:
\begin{equation}
  \label{eq:trunc-recipe-1}
  \ket{\psi} = \sum_{l\sigma r} A^{[\sigma]}_{l r} \ket{l}\ket{\sigma}\ket{r}.
\end{equation}
To construct a projection operator $P$ that projects out the high
energy components for this site, $\ket{\psi} \mapsto P \ket{\psi}$,
one may proceed as follows:

First, build a Krylov subspace of dimension $\dtrunc$ within
span$\{|l\rangle |\sigma\rangle | r\rangle \}$ and calculate the
matrix elements of $\hat \Hshift$ within it (no truncation necessary):
\begin{subequations}
\label{eq:trunc-recipe}
\begin{align}
  \label{eq:trunc-recipe-2}
  \ket{\tilde{i}} &= (\hat H')^{i-1}\ket{\psi},\qquad\qquad i = 1\ldots \dtrunc,\\
  \label{eq:trunc-recipe-3}
  \ket{\tilde{i}} &\mapsto \ket{i} \quad \textrm{orthonormalize via Gram-Schmidt,} \\
  \label{eq:trunc-recipe-4}
  (\hat H'_{K})_{ij} &= \bra{i}\hat H'_{K}\ket{j},\qquad\qquad 
  \hat H{'}^{K} \in \mathbb{C}^{\dtrunc\times \dtrunc}.
\end{align}
\end{subequations}
Next, fully diagonalize $\hat H'_{K}$ to obtain all eigenenergies
$\varepsilon'_\alpha$ and eigenvectors $\ket{e_\alpha}$
\begin{equation}
  \label{eq:trunc-recipe-5}
  \hat U^{\dagger} \hat H'_{K}  \hat U = \sum_{\alpha=1}^{\dtrunc} \ket{e_\alpha}
  \varepsilon'_\alpha\bra{e_\alpha}.
\end{equation}
Then construct the projection operator 
\begin{equation}
  \label{eq:trunc-recipe-6}
  P = \id - \sum_{\alpha: \varepsilon'_\alpha \ge \etrunc} \ket{e_\alpha}\!\bra{e_\alpha} 
\end{equation}
for a certain energy threshold $\etrunc$ and apply it:
\begin{equation}
  \label{eq:trunc-recipe-7}
  \ket{\psi} \mapsto P\ket{\psi}.
\end{equation}
Performing this procedure once for every site of the chain constitutes
a truncation sweep.  The state obtained after several truncation
sweeps, say $|t_n \rangle_\trunc$, is stripped from the
unwanted high-energy components of $\ket{t_{n}}$, as well as 
possible within a Krylov approximation.  After fitting and truncation
have been completed, the resulting (unnormalized) state $|t_n
\rangle_\trunc$ is renamed $|t_n \rangle$, used for calculating
$\mu_n$, and fed into the next recursion step.

To quantify the effects of energy truncation, we consider two
measures of how much $|t_n \rangle$ changes during
truncation. First, for a given truncation sweep, we define the
\emph{average truncated weight per site} (averaged over all sites)
by
\begin{equation}
  \label{eq:trunc-norm-per-sweep}
  N_{\trunc}^{\sweep} =
  \sqrt{\frac{1}{L}\sum_{k} \sum_{\alpha: \varepsilon'_\alpha \ge \etrunc} \left|\bracket{e_\alpha^{k}}{\psi}\right|^{2}} \,,
\end{equation}
where $\ket{e_\alpha^{k}}$ are the vectors constituting the projector
of \fref{eq:trunc-recipe-6} at site $k$.  Second, we define the
\emph{truncation-induced state change} by
\begin{eqnarray}
  \label{eq:Delta_trunc}
  \Delta_\trunc = \left\| |t_n \rangle_\trunc - |t_n \rangle \right\|^2 \, .
\end{eqnarray}
It measures changes in the state due to the intended truncation of
high energy weight, but also due to unavoidable numerical errors. In
our experience, neither of the truncation measures
$N_{\trunc}^{\sweep}$ and $\Delta_\trunc$ show clear signs of decay
when increasing the number of truncation sweeps, say $\ntrunc$ (see
\fref{fig:fidelity-compare}(c) below). This reflects the fact that
energy trunctation has the status of a precautionary measure, not a
variational procedure, and implies that there is no dynamic criterion
when to stop truncation sweeping. As a consequence, one has to analyze
how the accuracy of the results depends on $\ntrunc$ and optimize the
latter accordingly. This will be described in \fref{sec:CheMPS-ED-Comparison}
below.

The numerical costs for energy truncation are as follows: The cost for
the steps in \Eqs{eq:trunc-recipe} are
${\cal O}(\dtrunc^{2}m^{3}d^{2}D_{H})$, where $m$ is the MPS dimension,
$d$ the size of the local site basis and $D_{H}$ the matrix
product operator dimension of $\hat H'$.  The diagonalization of
$\hat H'{}^K$ is of ${\cal O}(\dtrunc^{3})$ where $\dtrunc$ is
theoretically bounded by $m^{2}d$.  In our experience, the purpose of
the energy truncation, which is solely to eliminate high-energy
contributions, is well accomplished already for a relatively small
Krylov subspace dimension of $\dtrunc=30 \ll m^{2}d$.

An overview of all the parameters relevant for \CheMPS is given in
\fref{tab:expansion-params}.  Where applicable, it also lists the
values that we found to be optimal.  A detailed error analysis,
tracing the effects of various choices for these parameters, will be
presented in \fref{sec:error-analysis}.

\section{Results: Heisenberg antiferromagnet}
\label{sec:heisenberg}

To illustrate the capabilities and power of the proposed \CheMPS
approach, this section presents results for the spin structure factor
of a one-dimensional spin-$\frac{1}{2}$ Heisenberg antiferromagnet
(HAFM) and compares them against results obtained from CV and tDMRG
approaches.

\subsection{Spin structure factor}
\label{sec:spin-structure}

We study the spin-$\frac{1}{2}$ HAFM for
a lattice of length $L$
\begin{equation}
  \label{eq:Heisenberg-ham}
  \hat H_{\rm HAFM} = J \sum_{j=1}^{L-1} \hat{ \vec{S}}_{j}\cdot \hat {\vec{S}}_{j+1} \,,
\end{equation}
where $\hat {\vec{S}}_{j}$ denotes the spin operator at site $j$.  We
choose $J=1$ as unit of energy throughout this section. This model
exhibits SU(2) symmetry, which has been
exploited\cite{McCullochGulacsi2002} in our calculations , accordingly all MPS dimensions noted for the HAFM are to be understood as number of SU(2) (representative) states being kept. To account
for the open boundary conditions we define spin wave operators as:
\begin{equation}
  \label{eq:spin-k}
  \hat{  \vec{S}}_{k} = \sqrt{\frac{2}{L+1}}\sum_{j=1}^{L} 
  \sin\frac{jk\pi}{L+1} \hat{\vec{S}}_{j} \,.
\end{equation}
The spin structure factor (spectral function) we are interested in is given by
\begin{equation}
  \label{eq:spin-spectral-function}
  S(k, \omega) = {\cal A}^{\vec{S}^\dag_{k} \cdot \vec{S}_{k}}(\omega) \,.
\end{equation}
It is known from exact solutions
\cite{CloizeauxPearson1962,FaddeevTakhtajan1981,MuellerBeckBonner1979,KarbachMuellerBougourziFledderjohannMuetter1997,Caux2006}
that the dominant part of the spin structure factor stems from two-spinon
contributions, bounded from below and above by
\begin{equation}
  \label{eq:spectrum-bound}
  \omega_{1} = \frac{\pi}{2}\abs{\sin k}\quad \text{and}\quad \omega_{2} = \pi\abs{\sin \frac{k}{2}} \, .
\end{equation}
Moreover, for an infinite system $S(k,\omega)$ is
known\cite{KarbachMuellerBougourziFledderjohannMuetter1997,Caux2006}
to diverge as
\begin{subequations}
\label{eq:diverging-peaks}
\begin{eqnarray}
  S(k,\omega) &\sim &  [\omega-\omega_{1}]^{-\frac{1}{2}} \sqrt{\ln[1/(\omega-\omega_{1})]} \; , \textrm{for }k\neq\pi , \qqph \\
  \label{eq:Sk=pi}
  S(\pi,\omega) &\sim & \omega^{-1} \sqrt{\ln(1/\omega)} \; ,
\end{eqnarray}
\end{subequations}
as $\omega$ approaches the lower threshold $\omega_1$ from
above. This divergence reflects the tendency towards staggered spin
order of the ground state of the Heisenberg antiferromagnet.  It poses
a severe challenge for numerics, which always deals with systems of
finite size, and hence will never yield a true divergence. Instead,
the divergence will be cut off at $\omega-\omega_1 \simeq 1/L$,
yielding a peak of finite height,
\begin{subequations}
\label{eq:max-peaks-height}
\begin{eqnarray}
  \max  S(k ,\omega) &\sim & [L \ln L]^{\frac{1}{2}} \; , \quad \textrm{for }k\neq\pi , \qph \\
  \max  S(\pi,\omega) &\sim &  L[ \ln L]^{\frac{1}{2}} \; . 
\end{eqnarray}
\end{subequations}
Thus, the best one can hope to achieve with numerics is to capture the
nature of the divergence as $\omega$ approaches $\omega_1$ before it
is cut off by finite size, or the scaling of the peak height with
system size.

\Eq{eq:spectrum-bound} gives a good guide for choosing $\Wast$. We
found the choices $\Wast=6.3 \simeq 2\pi$ and $\epsilont=0.025$ to
work well for all $k$ and have used them for all figures
(\ref{fig:spin-cv-compare} to \ref{fig:spin-L-compare}) of this
section. As consistency checks, we verified that the resulting $S(k,
\omega)$ is essentially independent of $\Wast$, and that it agrees
with a calculation that included the full many-body bandwidth $(\Wast = W)$.  

To have an accurate starting point for all calculations, we throughout
used a ground state obtained by standard DMRG with MPS dimension
$m=512$. From expansion order $n=1$ onwards, it turned out to be
sufficient to represent all \cheby vectors $\ket{t_{n}}$ using a
surprisingly small MPS dimension of $m=32$, or $m=64$ for some results
involving very large iteration number, as indicated in every
figure. (In retrospect, this implies that for the ground state, too, a
much smaller $m$ would have sufficed.)  We have verified that the
structure factor $S(k,\omega)$ is well converged w.r.t.\ $m$
nevertheless.  Detailed evidence for this claim will be presented
below. However, already at this stage it is worth remarking that
\emph{the ability of \CheMPS to get good results with comparatively
  small $m$-values is perhaps the single most striking conclusion of
  our work}. This will be discussed in detail below.

\subsection{Comparison to CV}
\label{sec:comparison-cv}

\begin{figure}[tb]
  \centering
  \includegraphics[width=\linewidth]{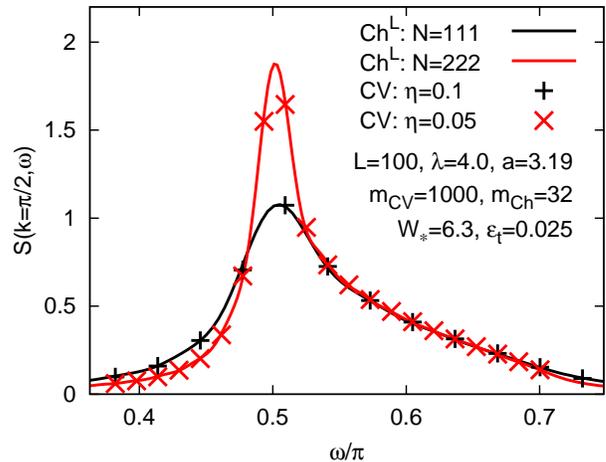}
  \caption{(Color online) Comparison of \CheMPS vs.\ correction vector (CV)
    calculations of $S(k=\pi/2, \omega)$ for a Heisenberg chain:
    Lines show \cheby results reconstructed for $N=118$ and 236
    using Lorentz damping, \fref{eq:app-cheby-13}, with
    $\lambda = 4.0$; symbols show CV results, obtained using
    broadening parameters of $\eta = 0.1$ and 0.05. We expect and
    indeed find good agreement between lines and symbols, since both
    Lorentz damping and the correction vector method in effect
    broaden the spectral function by Lorentzians, whose widths we
    equated by choosing $a \lambda /N = \eta$ (with $a = 3.19$), see
    \Eq{eq:cv-compare-N}, and also \fref{fig:kernel-delta}. Since
    $m_{\rm Ch} \ll m_{\rm CV}$, the numerical cost of obtaining an
    \emph{entire} curve via Chebychev is dramatically cheaper than
    calculating a \emph{single} point via CV, as discussed in the
    text.}
  \label{fig:spin-cv-compare}
\end{figure}
We begin our discussion of \CheMPS results by comparing them to those
of CV calculations, which are known to be very accurate, though also
computationally expensive. The CV method involves a broadening
parameter $\eta$ and broadens $\delta$-functions into Lorentzian peaks
of width $\eta$. This can be mimicked with \CheMPS by using Lorentz
damping (with $\lambda = 4.0$), since this also produces Lorentzian
broadening, representing a $\delta$-function $\delta (\omega' -
\bar \omega')$ by a near-Lorentzian peak, albeit with a
frequency-dependent width, $\eta'_{N, \bar \omega'} = \sqrt{1 - \bar
  \omega'^2} \, \lambda/N$ (see Sec.~\ref{sec:cheby-basics} and
\fref{fig:kernel-delta}).  To compare \CheMPS results with CV
results at given $\eta$, we thus identify $\eta = a \eta'_{N, \bar
  \omega'}$, where $a$ is the scaling factor from
\Eqs{eq:mapping-omega-H} and $\bar \omega'$ is taken to be the
rescaled and shifted version of the frequency $\omega_{\rm max}$ at
which the peak reaches its maximum. Thus we set the expansion order
used for reconstruction to
\begin{equation}
  \label{eq:cv-compare-N}
  N = \frac{4a}{\eta }{\sqrt{1 - (\omega_{\rm max}/a - W')^2}} \, .
\end{equation}
Figure~\ref{fig:spin-cv-compare} shows such a comparison for the
structure factor $S(\pi/2,\omega)$ of a $L=100$ Heisenberg
chain. We used two choices of $\eta$ that are large enough to avoid
finite-size effects, namely $\eta= 0.1$ and 0.05, and set
$\omega_{\rm max} = \pi/2$ (cf.\ $\omega_1$ of
\Eq{eq:spectrum-bound}).  We used MPS dimensions of $m_\textrm{CV}
=1000$ or $m_{\rm Ch} = 32$ for CV or \CheMPS calculations,
respectively.  (Our choice for $m_{\rm CV}$ aimed for achieving highly
accurate CV results; for $\eta=0.05$ this required $m_{\rm CV}=1000$,
but for $\eta=0.1$, a slightly smaller value for $m_{\rm CV}$ would
have sufficed.) We find excellent agreement between the two approaches
without adjusting any free parameter, since $N$ is fixed by
\fref{eq:cv-compare-N}. For example, for $\eta = 0.05$, $N=255$, the
relative error is less than 3 \% for all $\omega$.

Since this level of agreement is obtained using $m_{\rm Ch} \ll m_{\rm CV}$,
we conclude that \CheMPS with Lorentz damping gives results
whose accuracy is comparable to those of CV, \emph{at dramatically reduced
numerical cost}. Indeed, for $\eta = 0.05$ the calculation of the
entire \CheMPS spectral function was 25 times faster than that of a
single CV data point.

\subsection{Finite-size effects} 
\label{sec:finite-size}

Let us now analyse the role of finite system size. To
this end, it is of course important to understand broadening effects
in detail.  The fact that \CheMPS offers simple and
systematic control of broadening via the choice of the expansion
order (and damping factors), as will be illustrated below, is very
convenient and may regarded as one of its main advantages.

\begin{figure*}[tb]
  \centering
  \includegraphics[width=\linewidth]{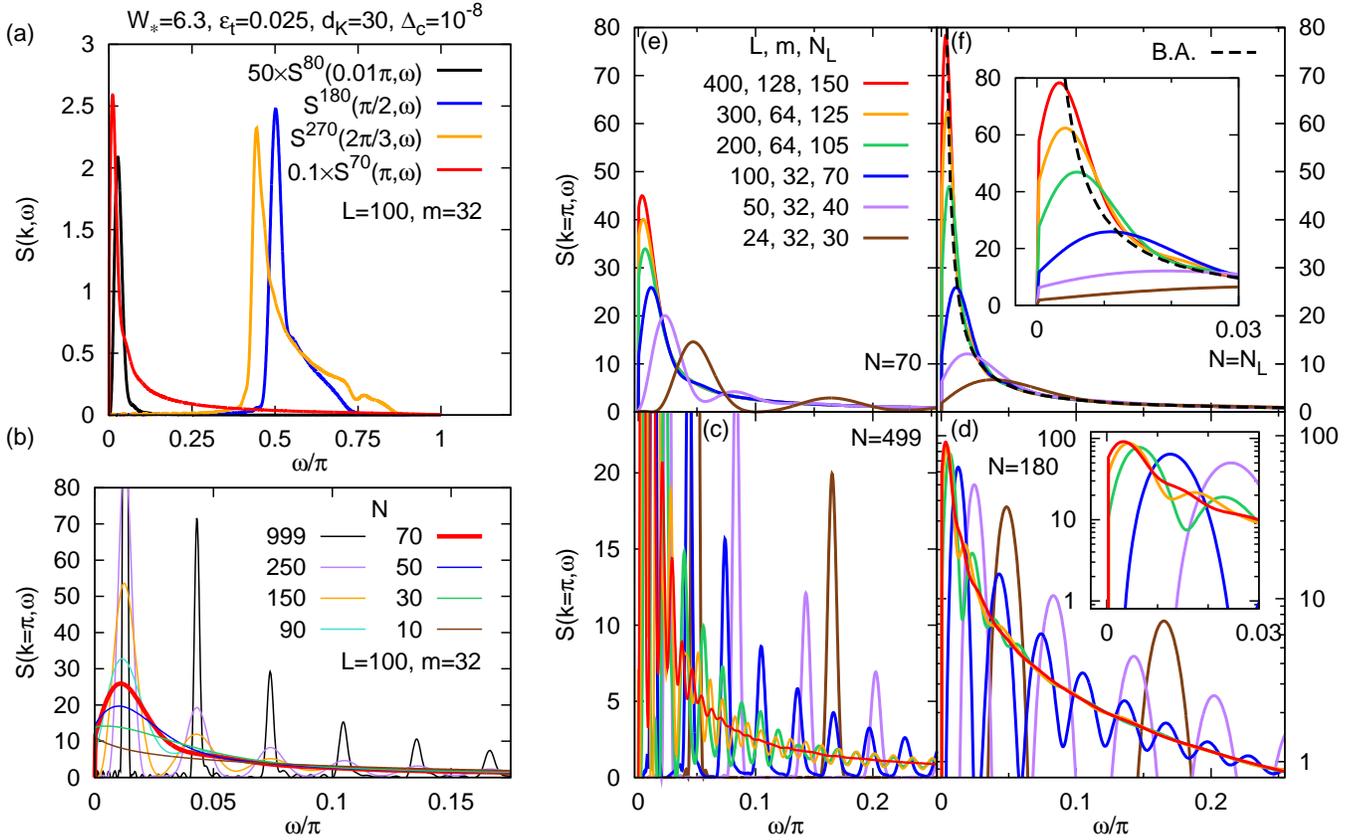}
  \caption{(Color online) Spin structure factors for a Heisenberg
    chain, reconstructed using Jackson damping.  (a) $S(k,\omega)$ for
    four choices of momentum $k$, for a chain of length $L=100$. Each
    spectrum was reconstructed using \obroadening, i.e.\ by choosing
    the largest expansion order, say $N_L$ (indicated by superscripts),
    that does not yet resolve finite-size effects. (b-f) Finite-size
    analysis of $S(\pi,\omega)$. (b) To determine $N_L$ for given $L$
    (here 100), several different expansion orders $N$ are considered.
    Increasing $N$ reduces the effective broadening $\etaN
    \simeq\mathcal{O}( \Wast/N)$ until finite-size subpeaks appear for
    $N > N_L$ (here $N_L= 70$, bold red curve). (c) Evolution of the
    finite-size structure with $L$, revealed by fixing $N$ large enough
    (here = 499) to resolve the first few dominant subpeaks of all
    curves. There are $L$ dominant subpeaks (not all shown here) within
    the spectral bandwidth, with average spacing $\omega_L \sim 1/L$.
    (d) Same as (c), but plotted on a semi-log scale, and with somewhat
    smaller $N$ (here = 180), chosen to be somewhat larger than the
    optimal broadening $N_L$ for the largest $L$ (here $N_{300}=
    125$). As $L$ increases and $\omega_L$ decreases, the subpeaks
    coalesce toward the intrinsic lineshape $S^\infty (k,\omega)$.  (e)
    When $L$ is increased at \emph{fixed} $N$ (here 70), finite-size
    effects disappear once $\omega_L$ drops below the effective
    broadening $\etaN$, resulting in a smooth spectral function.  (f)
    In contrast, when $L$ is increased while using \emph{optimal
      broadening}, $N=N_L$ (i.e.\ $\eta_N$ just above $\omega_L$), none
    of the curves show finite-size effects, and the resulting main peak
    is sharper than in (e).  In both (e) and (f), the peak height shows
    no indications of converging with $L$, reflecting the fact that the
    true peak shape involves an $\omega^{-1}[ \ln \omega]^{-1/2}$
    divergence. Moreover, the \CheMPS curves in (f) show signs of
    overbroadening when compared to the exact Bethe Ansatz result
    (dashed), from Ref.~\onlinecite{Caux2006}.}
  \label{fig:finite-size-analysis}
\end{figure*}

Figure~\ref{fig:finite-size-analysis}(a) shows \CheMPS results for the
spin structure factors $S(k,\omega)$ of four different momenta $k$,
calculated for $L=100$ using Jackson damping. They were reconstructed
using the largest expansion order, say $N_L$, that does not yet
resolve finite-size effects, a choice that will be called
\emph{\obroadening.} Each curve shows a dominant peak, and we are
interested in finding its intrinsic shape $S^\infty(k,\omega)$
in the continuum limit of an infinitely long chain ($L\to \infty$).
Thus, the following general question arises: under what conditions
will a spectrum calculated for finite system size $L$ and
reconstructed with finite expansion order $N$, say $S_N^L(
\omega)$, correctly reproduce the desired continuum spectrum
$S^\infty (\omega)$? The general answer, of course, is that the
optimally broadened spectrum should have converged as function of $L$,
i.e.\ the shape of $S_{N_L}^L( \omega)$ should not change
upon increasing $L$.  However, for a spectrum with an intrinsic
divergence, such as \Eq{eq:diverging-peaks}, the peak's height will
never saturate with $L$; at best one can hope to observe
$L$-convergence of the shape of its tail, and the proper scaling of
its height (\Eq{eq:max-peaks-height}).

To illustrate the nature of finite-size effects and the role of $N$ in
revealing or hiding them, \fref{fig:finite-size-analysis}(b) shows
$S(\pi, \omega)$ for $L=100$ and several values of $N$, both smaller
and larger than $N_L$. As $N$ is increased and the effective
broadening $\etaN \simeq \mathcal{O}(W_\ast/N)$ decreases, the main
peak of the initially very broad and smooth spectral function becomes
sharper. Optimal broadening in \fref{fig:finite-size-analysis}(b)
corresponds to $N_L \simeq 70$, beyond which additional ``wiggles''
emerge. These develop, with beautifully uniform resolution, into
\emph{dominant subpeaks} as $N$ is increased further. The
discrete subpeaks reflect the quantized energies of spin-wave
excitations in a finite system.  With sufficiently high resolution
($N=999$ in \fref{fig:finite-size-analysis}(b)) numerous additional
minor subpeaks emerge, but their weight is very small compared to
that of the dominant subpeaks. This fact is important, since it
implies that the structure factor of a finite-size system is
exhausted almost fully by the set of dominant subpeaks, with very
small intrinsic widths. 

We have checked that there are $\mathcal{O}(L)$ dominant subpeaks
within the spectral bandwidth of $S(k,\omega)$.  Correspondingly, the
average spacing between dominant subpeaks, to be called the
\emph{finite-size energy scale} $\omega_L$, is proportional to
$\frac{1}{L}$ (\fref{fig:finite-size-analysis}(c,d)).  The weight of
each subpeak decreases similarly, ensuring that the total weight in a
given frequency interval converges as $L\to \infty$.  The inverse
subpeak spacing $\hbar/ \omega_L$ corresponds to the Heisenberg time,
i.e.\ the time within which a spin wave packet propagates the length
of the system.

Figures~\ref{fig:finite-size-analysis}(e,f) illustrate two slightly
different broadening strategies.  In
\fref{fig:finite-size-analysis}(e), $L$ is increased for fixed $N$:
the distinct subpeaks increasingly overlap, resulting in a smooth
spectral function once $\omega_L$ drops below $\etaN$. In
\fref{fig:finite-size-analysis}(f) optimal broadening is used
($\etaN$ just larger than $\omega_L$: now no subpeaks are visible,
and the $L$-evolution of the main peak is revealed with better
resolution.

\begin{figure}[tb]
  \centering
  \includegraphics[width=\linewidth]{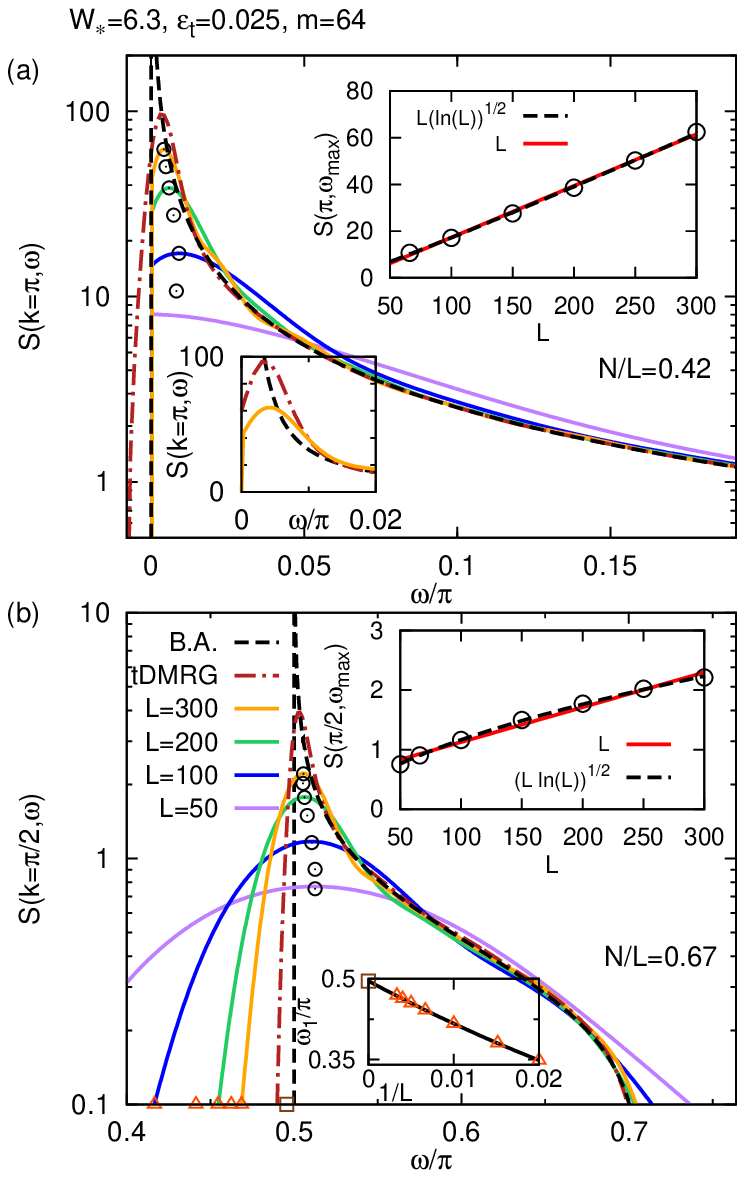}
  \caption{(Color online) Comparison of \CheMPS, Bethe Ansatz and
    tDMRG + prediction, for HAFM structure factors with (a) $k=\pi$
    and (b) $k=\pi/2$. Dashed lines: Bethe Ansatz results for
    $L=\infty$,  from Ref.~\onlinecite{Caux2006}. Dashed-dotted
    lines: tDMRG results,  from
    Ref.~\onlinecite{BarthelSchollwoeckWhite2009}. Other lines show
    \CheMPS results for $L=50, 100, 200, 300$, reconstructed using a
    fixed ratio of $N/L$, namely 0.42 for (a) and 0.67 for (b).
    Circles mark \cheby peak maxima,
    also for $L=66, 150$ and 250, for which no curves are shown.  The
    lower inset in (a) zooms into the peak region using a linear scale,
    illustrating overbroadening.  The upper insets of (a,b) show
    the peak heights vs.\ $L$ (circles), together with a fit to the
    Bethe-Ansatz expectation from \Eq{eq:diverging-peaks}
    (dashed line) or to a straight line (solid line). In
    (b), triangles mark the lower threshold frequencies for which
    $S(\pi/2,\omega)$ equals a fixed, small value, arbitrarily chosen
    as 0.1. Their $1/L\to 0$ extrapolation, shown in lower inset,
    gives an estimate for the lower threshold frequency, namely
    $\omega_1/\pi = 0.496$ (marked by a square); the exact value is
    $1/2$.}
  \label{fig:spin-L-compare}
\end{figure}
In both Figs.~\ref{fig:finite-size-analysis}(e) and
\ref{fig:finite-size-analysis}(f), the peak height shows no
indications of converging with increasing $L$.  (The same is true for
the data of \fref{fig:finite-size-analysis}(a).) This reflects the
intrinsic divergence of the peak height expected from
\Eq{eq:diverging-peaks}. Figures~\ref{fig:spin-L-compare}(a) and
\ref{fig:spin-L-compare}(b) contain a quantitative analysis of this
divergence, for $S(\pi,\omega)$ and $S(\pi/2,\omega)$, respectively.
The shape of the divergencies for an infinite system are shown by the
thick solid lines, representing exact Bethe Ansatz results from
Ref.~\onlinecite{Caux2006}. Thin dashed lines show results from tDMRG
from Ref.~\onlinecite{BarthelSchollwoeckWhite2009} for $L=100$, and
thin solid lines \CheMPS results for several system sizes between
$L=50$ and 300. For \CheMPS spectral reconstruction, we determined the
expansion order $N_{300}$ that ensures optimal broadening for $L=300$,
and used a fixed ratio of $N/L = N_{300}/300$ for all curves (namely
0.42 or 0.67 for Figs.~\ref{fig:spin-cv-compare}(a,b),
respectively). \CheMPS (for $L=300$) and tDMRG reproduce the peak's
tail and flank well, but clearly and expectedly are unable to produce
a true divergence at the lower threshold frequency. Nevertheless, the
insets show that the manner in which the \CheMPS peak heights increase
with $L$ is indeed consistent with \Eq{eq:diverging-peaks}. (For
the limited range of available system sizes, however, a reliable
distinction between $L [\ln(L)]^{1/2}$, $[L \ln(L)]^{1/2}$ or $L$
behavior is not possible.)

It is also possible to determine the lower threshold frequency
$\omega_1$ rather accurately from the \CheMPS results by doing an
$1/L$ extrapolation. We illustrate this in
\fref{fig:spin-L-compare}(b) by extrapolating the frequencies at which
$S(\pi/2,\omega)=0.1$ (triangles). Since the data exhibit a slight
curvature when plotted against $1/L$ (see lower inset of
\fref{fig:spin-L-compare}(b)), they were fitted using a second order
polynomial in $1/L$. Extrapolating the fit to $1/L = 0$ yields
$\omega_1 = 0.496 \pi$ (marked by a square), in good agreement
with the prediction $\omega_1 = \pi/2$ from \Eq{eq:spectrum-bound}.

\subsection{Discrete representation of spectral function}
\label{sec:discrete-spectrum}

In both \fref{fig:finite-size-analysis}(f)  and
\fref{fig:spin-L-compare}, the right flank of the peak still bears 
signatures of overbroadening: the curve for a given $L$
lies above those for larger $L$ (before bending
over towards its peak), and all curves lie significantly
above the exact Bethe Ansatz curve (dashed line).
One way of reducing this broadening would be to
simply increase $L$, but this is numerically costly. Clearly,
alternative strategies for reducing finite-size effects would be
desirable. One such scheme, involving linear prediction in the time domain,
will be discussed in the next subsection. Here we present
another,  which exploits the ability of \CheMPS to accurately
resolve finite-size peaks. 

The origin of overbroadening is clear: when neighboring subpeaks are
broadened enough to overlap, weight is inevitably transfered from
large peaks to smaller peaks. This effect is negligible only in the
limit $L\to \infty$, where the subpeak spacing becomes negligible.  To
avoid overbroadening for a finite-$L$ system, one thus has to
analyse spectra for which $N$ is large enough that subpeaks do
\emph{not} overlap significantly, such as that shown in
\fref{fig:finite-size-analysis}(b).

To be concrete, let us represent the true, discrete spectrum of a
system of size $L$ by a sum of peaks, enumerated by a counting index
$\alpha$, with position $\Omega_\alpha$, width $\bar \eta_\alpha$, weight
$W_\alpha$ and Gaussian shape $K^J$ (cf. \Eq{eq:kernels-J}):
\begin{eqnarray}
  \label{eq:fit-CheMPS-delta-peaks}
  S^{L} (k,\omega) \simeq \sum_{\alpha} W_\alpha K^J_{\bar \eta_\alpha} (\omega - \Omega_\alpha) \; .
\end{eqnarray}
Its \cheby reconstruction with Jackson damping, say $S_N^L(k,
\omega)$, will have the same form, except that the peaks will be
broadened to have widths, $\eta_\alpha = (\bar \eta_\alpha^2 +
\eta_{N,\alpha}^2)^{1/2}$, as explained before
\Eq{eq:adding-Gaussian-widths}. If $N$ is large enough, the broadened
peaks will still be clearly separated (as for $N = 999$ or 250 in
\fref{fig:finite-size-analysis}). By fitting each peak (separately,
one by one) to a Gaussian, one can determine its position
$\Omega_\alpha$, weight $W_\alpha$ and effective width $\eta_\alpha$,
and deduce the intrinsic width via $\bar \eta_\alpha = (\eta_\alpha^2
- \eta_{N,\alpha}^2)^{1/2}$. We find (not shown) that the intrinsic
width grows with increasing frequency $\Omega_\alpha$. This implies,
not unexpectedly, that higher-lying spin-wave excitations have shorter
life-times. However, it also implies that higher-lying peaks
eventually start to overlap, so that the analysis to be described
below is feasible only for a limited number of low-lying peaks.

The discrete peaks suggest a natural partitioning of the frequency
spectrum into intervals $I_\alpha$: each contains one peak of weight
$W_\alpha$ at position $\Omega_\alpha$, extends halfway to the next
peaks at $\Omega_{\alpha \pm 1}$ on either side, and has width
$\Delta_\alpha = (\Omega_{\alpha+1} - \Omega_{\alpha-1})/2$.  The
first interval above the lower spectral threshold ($\omega_1$) is
defined slightly differently: $I_1$ has lower bound $\omega_1$ and
width $\Delta_1 = (\Omega_1 + \Omega_2)/2 - \omega_1$.

Now, to produce a smooth curve devoid of finite-size effects, the
subpeaks must be broadened until they overlap substantially. However,
if the weights in two neighboring intervals differ, say $W_\alpha >
W_{\alpha+1}$, such broadening inevitably transfers weight from
interval $I_\alpha$ to $I_{\alpha +1}$, resulting in overbroadening.

Such overbroadening can be avoided by constructing a \emph{discrete}
representation of the spectral function, $S_{\rm dis}(k, \Omega_\alpha)$,
defined by the set of coordinates
\begin{eqnarray}
  \label{eq:discrete-spectral-function}
  \{(\Omega_\alpha, S_\alpha)\} ,  \quad \textrm{with} \quad   S_\alpha  =
  S_{\rm dis} (k, \Omega_\alpha) =  W_\alpha / \Delta_\alpha \; . \qph
\end{eqnarray}
The identification of $S_\alpha$ with 
$W_\alpha/\Delta_\alpha$ follows from applying the definition of 
a spectral function, namely spectral weight per unit frequency
interval, to the interval $I_\alpha$.

\begin{figure}[tb]
  \centering
  \includegraphics[width=\linewidth]{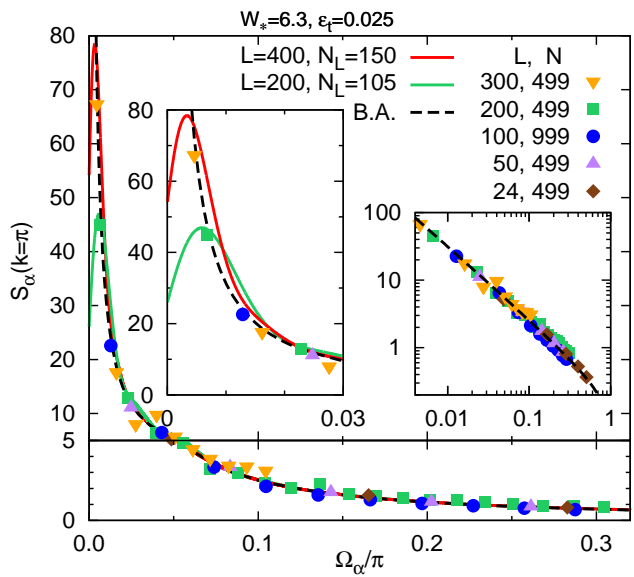}
  \caption{(Color online) Discrete representation
    (\Eq{eq:discrete-spectral-function}) of the structure factor
    $S_\alpha = S_{\rm dis} (\pi,\Omega_\alpha)$ (symbols), for five
    different system sizes. (The lower panel uses an enlarged vertical
    scale, to zoom in on the tail region.)  For comparison, the Bethe Ansatz
    result (dashed line) and two optimally broadened spectra, for
    $L=400$ and 200 (solid lines), are also shown.  Left inset: zoom
    to low frequencies, showing that the discrete data completely
    avoids overbroadening, in contrast to the optimally broadened
    spectra. Right inset: log-log version of main plot. The frequency
    range does not extend low enough to be able to uncover the pure
    asymptotic predicted by \Eq{eq:Sk=pi}.}
  \label{fig:discrete-spectrum}
\end{figure}

Figure~\ref{fig:discrete-spectrum} shows the resulting discrete data
points for four different system sizes. Remarkably, they all fall onto
the same curve, which agrees well with the Bethe Ansatz result (dashed
line). In particular, the first two or three data points for each $L$
lie right on top of the Bethe Ansatz curve (dashed), see
\fref{fig:discrete-spectrum}, left inset, beautifully mapping out the
true shape of the spectral function down to the lowest discrete
excitation frequency $\Omega_\alpha$ that exists for that
$L$. Evidently, the discrete spectral function is completely
\emph{free from broadening artifacts}, in marked contrast to the
optimally broadened curves shown for $L=200$ and 400 (solid lines)
(compare also \fref{fig:finite-size-analysis}(f)). This advantage
comes at the price of specifying the spectral function only at
discrete points, not via a continuous curve. However, for a system of
finite size, such discreteness is fundamentally unavoidable. The good
news is that the continuum curve $S^\infty_N (k ,\omega)$ is evidently
well mimicked by the discrete representation $\{(\Omega_\alpha,
S_\alpha)\}$, and that \CheMPS allows the latter to be determined in a
straightforward fashion for system sizes well beyond what can be
done with exact diagonalization. We are not aware of any other
numerical many-body method capable of doing so for system sizes as
large as those considered here.

For larger frequencies the scatter of the discrete data w.r.t.\ the
Bethe Ansatz curve increases, reflecting the fact that subpeaks begin
to overlap there, making the extraction of discrete data increasingly
difficult. However, this is not a serious concern, since
in this frequency regime optimal broadening is able
to produce smooth spectra in good agreement with Bethe
Ansatz anyway.

To conclude this subsection, let us summarize the two main results of
our finite-size analysis.  The first concerns physics: for a chain of
finite chain of $L$ sites, the structure factor is dominated by a set
of $\mathcal{O}(L)$ sharp subpeaks, whose spacing and weight scale as
$1/L$. The second concerns methodology: \CheMPS very conveniently
allows this structure to be revealed or hidden, by simply choosing $N$
appropriately. Moreover, it can exploit information on the
positions and weights of the discrete subpeaks to largely eliminate
broadening artefacts.

\subsection{Comparison of t\CheMPS to tDMRG} 
\label{sec:time-dependent-chemps}

Another possible scheme for reducing finite-size effects is to work
in the \emph{time domain} using linear prediction, as shown in
Ref.~\onlinecite{BarthelSchollwoeckWhite2009} for the HAFM.  The
idea is to calculate the Fourier transform of $S(k,\omega)$, namely
\begin{equation}
  \label{eq:Sxt-FT}
  S(k,t) = \sum_{x=1}^{L} e^{ik(x-x')} \expect{\hat 
    {\vec{S}}_{x}(t) \hat{\vec{S}}_{x'}(0)} \, ,
\end{equation}
with $x'$ chosen near the middle of the chain, and $t$ chosen small
enough that the spin excitation created at $x'$ does not reach the
edge of the system within $t$.  The function $S(k,t)$ thus obtained
will contain only weak finite-size effects. It is then extrapolated to
larger times via linear prediction techniques,
\cite{PressTeukolskyVetterlingFlannery2007,PereiraWhiteAffleck2008,WhiteAffleck2008,BarthelSchollwoeckWhite2009}
exploiting the fact that momentum excitations typically exhibit damped
harmonic dynamics, whose time-dependence can be extrapolated quite
accurately. Since the extrapolated function extends to very large
times, its Fourier transform yields good spectral resolution at low
frequencies\cite{PressTeukolskyVetterlingFlannery2007} (with an
accuracy that depends on that achieved during linear prediction).

In Ref.~\onlinecite{BarthelSchollwoeckWhite2009} the input correlator
needed for linear prediction, $ S(k,t)$, was calculated using
tDMRG. (Two examples of the resulting spectra are included in our
Fig.~\ref{fig:spin-L-compare}.)  We note that $ S(k,t)$ can also be
calculated using \CheMPS in the time-domain, to be called
t\CheMPS. Indeed the numerical cost for calculating $ S(k,t)$ by
evaluating the requisite correlators
$\expect{\hat{\vec{S}}_{x}(t)\hat{\vec{S}}_{x'}(0)}$ via
\Eq{eq:cheby-Gt} is essentially the same as calculating its Fourier
transform $S(k,\omega)$ via \Eq{eq:spectral-exp-A}, since the
corresponding \cheby moments $\mu_n$ can be calculated using the same
recursion scheme. In fact, if one defines $ \hat{\vec{S}}_{k}$ in
\Eq{eq:spin-k} using a pure exponential $e^{i k j}$ instead of a sin
function, the \cheby moments needed for $ S(k,t)$ are simply linear
combinations of those of $ S(k,\omega)$.

\begin{figure}[tb]
  \centering
  \includegraphics[width=\linewidth]{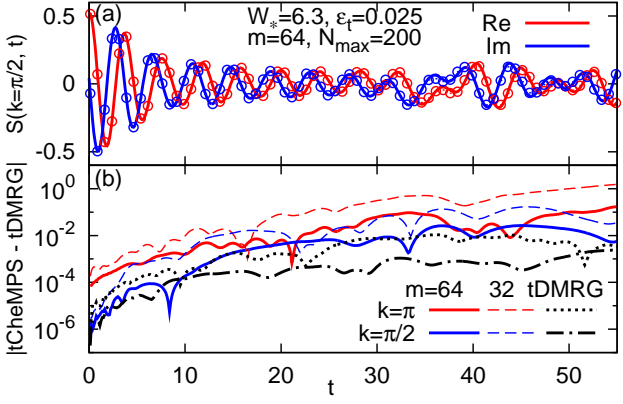}
  \caption{(Color online) (a) Time dependence of $S(\pi/2,t)$,
    calculated with t\CheMPS (lines) and tDMRG (symbols). Solid and
    dashed lines show, respectively, the real and imaginary parts of
    $S$. (b) The differences between t\CheMPS and tDMRG (with a
    specified truncation error of $10^{-6}$) of $S(k,t)$ for two
    values of $k$, and two values of $m$ (dashed/solid). To estimate
    the accuracy of tDMRG, we also show (dashed-dotted) the
    differences between two tDMRG calculations performed with
    different truncation error thresholds, namely $10^{-5}$ and
    $10^{-6}$, requiring up to $m=75$ or 125 states, respectively.}
  \label{fig:time-skt}
\end{figure}

To gauge the accuracy of t\CheMPS, we have calculated $S(\pi/2,t)$
using both t\CheMPS and tDMRG. Figure~\ref{fig:time-skt}(a) compares
the results, and \fref{fig:time-skt}(b) characterizes the differences.
We view the tDMRG results as benchmark, because for the times of
interest, we have checked them to be well converged (with errors
$\lesssim 10^{-3}$ for $t < 50$, see \fref{fig:time-skt}(b),
dashed-dotted line).  As expected, the agreement between t\CheMPS
and tDMRG is better for larger $m$. The differences are very small,
but grow with time, from being (for $m= 64$) below $10^{-3}$ for $t
\lesssim 10$ to around $10^{-2}$ for $t \simeq 30$, beyond which
finite-size effects start to appear.  

More generally, the results of \fref{fig:time-skt} illustrate that
\CheMPS offers a viable route to time evolution for situations where
extreme accuracy is not required. Further comments on this prospect
are included in the outlook, Sec.~\ref{sec:conclusion}.

\section{Error analysis}
\label{sec:error-analysis}

The convergence properties of a \cheby expansion are mathematically
well controlled and understood (see \Eq{eq:uniform-convergence}),
provided that the \cheby moments $\mu_n$ are known precisely. Their
evaluation via \CheMPS, however, introduces various sources of
numerical errors. This section is devoted to an analysis of these
errors.  In particular, we seek to determine appropriate choices for
the control parameters associated with the various \CheMPS tasks
listed in Table~\ref{tab:expansion-params}. We perform this
analysis mostly for a resonant level model (RLM), describing three
local levels coupled to a fermionic bath.  This model is introduced
and discussed in App.~\ref{sec:resonant-level-model}, which,
for the sake of completeness, also includes \CheMPS expansions of
the corresponding spectral functions. However, the details presented
there are not needed for the following discussion.

For the RLM, on the one hand, the \CheMPS evaluation of the $\mu_n$
is feasible to arbitrarily high orders, and on the other, exact
diagonalization (to be denoted by sub- or superscript ED) of the
single-particle Hamiltonian allows both the spectral function and
the \cheby moments $\mu_n$ to be found exactly. We use the
RLM-parameters specified in App.~\ref{sec:resonant-level-model}
throughout and focus mainly on the properties of one of its
correlators, $\mathcal{A}^-_{11}$ (without displaying corresponding
sub- and superscripts), which is defined in \Eq{eq:rlm-A-sum} and
whose behavior is representative for that of
$\mathcal{A}^{\pm}_{ij}$.

\subsection{Definition of error measures}
\label{sec:error-measures}

\begin{figure*}[tb]
  \centering
  \includegraphics[width=\linewidth]{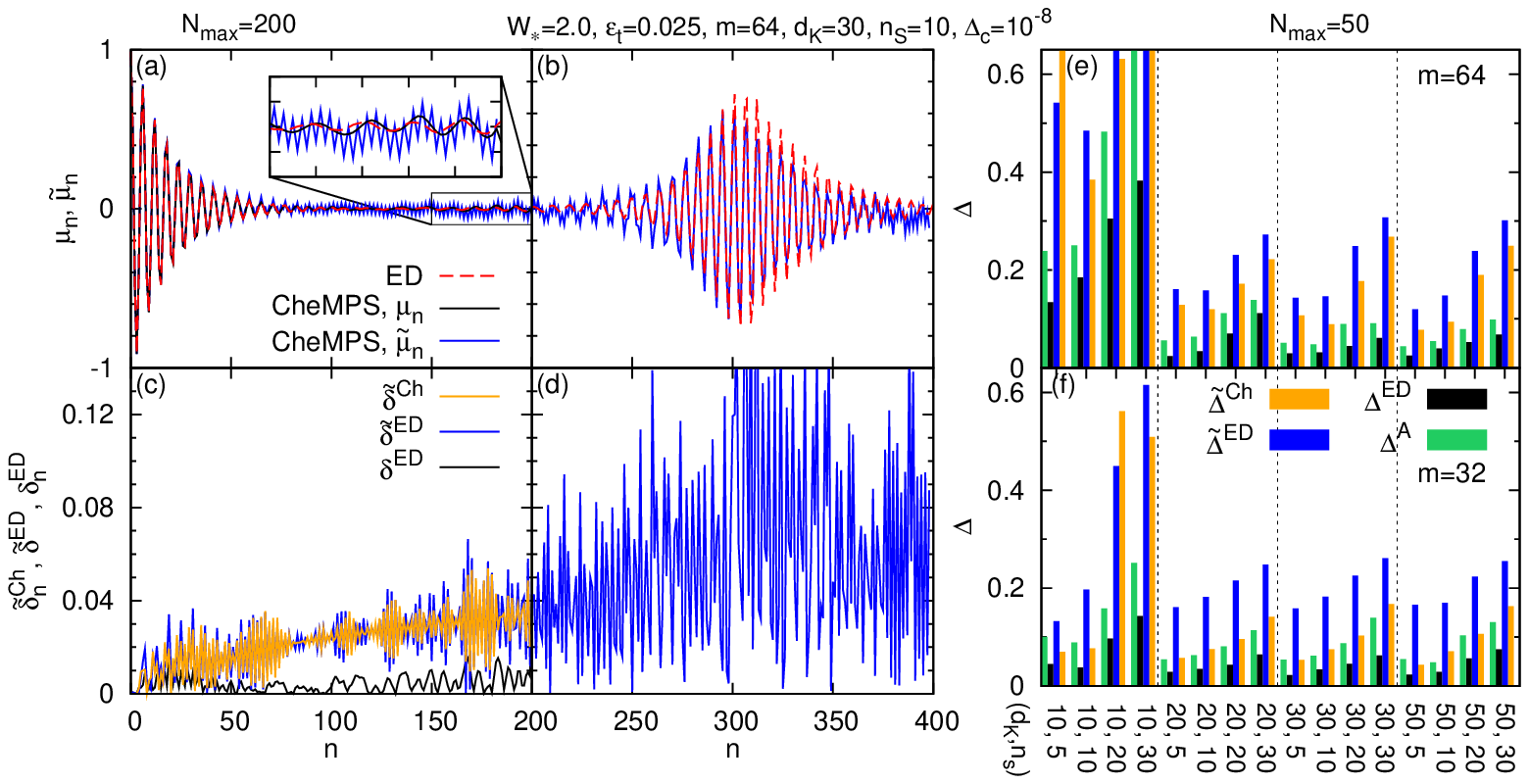}
  \caption[Expansion moments for the negative branch of $A_{11}$]{
    Comparison of \CheMPS and ED results for \cheby moments of the RLM
    spectral function $A_{11}^-$. (a,b) show $\mu_n$- and $\tilde
    \mu_n$ moments (\Eqs{eq:dmrg-recipe-1}, (\ref{eq:dmrg-recipe-4}))
    and (c,d) the $n$-dependent error
    measures $\delta^\ED_n$, $\tilde \delta^\ED_n$ and $\tilde
    \delta^\Ch_n$ (Eqs.~\ref{eq:def-deltas}), plotted in (a,c) for
    $n < \Nmax = 200$ and in (b,d) for $\Nmax \le n < 2 \Nmax$.  In
    (b), the increase in moment magnitude starting around $n \simeq
    250$ marks the onset of resolving finite-size structure in the
    spectral function.  (e,f) show the cumulative error measures
    $ \Delta^{\ED}$, $\tilde  \Delta^{\ED}$, $\tilde  \Delta^\Ch$
    (\Eqs{eq:delta-exact-int})  and $\Delta^A$ (\Eq{eq:delta-spectral-int})
    for various combinations of the MPS dimension $m$, the number 
    of energy truncation sweeps $\ntrunc$ and the Krylov subspace 
    dimension $\dtrunc$.}
  \label{fig:rlm-coeffs-illustrate}
\end{figure*}
We will analyse both $\mu_n$- and $\tilde \mu_n$-moments,
calculated from Eqs.~(\ref{eq:dmrg-recipe-1}) and
(\ref{eq:dmrg-recipe-4}), respectively. The differences between
\CheMPS and ED can be quantified by the error measures
\begin{subequations}
  \label{eq:def-deltas}
  \begin{eqnarray}
    \label{eq:def-delta-exact}
    \delta^\ED_n & = & \left|\mu_{n}^{\rm \CheMPS} - \mu_{n}^\ED \right| \,, \quad n < \Nmax \,,\\
    \label{eq:def-delta-exact-tilde}
    \tilde \delta^\ED_n & = & \left|\tilde \mu_{n}^{\rm \CheMPS} - \mu_{n}^\ED \right| \, , \quad n <  2 \Nmax  \,.
  \end{eqnarray}
  Moreover, to characterize the accuracy of \CheMPS moments
  without referring to exact results, we also consider
  \begin{eqnarray}
    \label{eq:def-delta-estimate}
    \tilde \delta^\Ch_n & = & \left|\tilde \mu_{n}^{\rm \CheMPS} - \mu_{n}^{\rm \CheMPS} \right| \,, 
    \quad n < \Nmax  \,.
  \end{eqnarray}
\end{subequations}

We will also use cumulative versions of these, namely
\begin{subequations}
  \label{eq:delta-exact-int}
  \begin{align}
    \label{eq:delta-exact-int-a}
    \Delta^{\ED}&= \sqrt{\sum_{n=0}^{\Nmax-1} \left(\delta_n^\ED \right)^2} \;, \\
    \label{eq:delta-exact-int-b}
    \tilde\Delta^\ED &= \sqrt{{\Delta^{\ED}}^{2} + \sum_{n=\Nmax}^{2\Nmax-1} (\tilde \delta_n^\ED)^2 } \;,\\
    \label{eq:delta-exact-int-c}
    \tilde \Delta^\Ch &= \sqrt{\sum_{n=0}^{\Nmax-1} (\tilde\delta_n^{\Ch})^2 } \;.
  \end{align}
\end{subequations}
Furthermore, we also introduce an integrated error measure for
undamped spectral functions (using Jackson damping would yield qualitatively similar error measures):
\begin{equation}
  \label{eq:delta-spectral-int}
  \Delta^{\rm A} = \sqrt{ \int_{0}^{\Wast} d\omega \left|\mathcal{A}^{2\Nmax}(\pm \omega) - \mathcal{A}^{\infty}(\pm \omega)\right|^{2} }\, .
\end{equation}
Here we use $\pm$ for $\mathcal{A^\pm}(\omega)$ spectra proportional
to $\theta (\pm \omega)$ (see \Eq{eq:rlm-A-sum}), and employ
$\mu_n$-moments for $n < \Nmax$ and $\tilde \mu_n$-moments for $ \Nmax
\le n < 2\Nmax$ during spectral reconstruction. (Note that $
\tilde\Delta^\ED$ of \Eq{eq:delta-exact-int-b} was constructed to
reflect this combination of $\mu_n$ and $\tilde \mu_n$.)

\subsection{Comparison of \CheMPS and ED moments}
\label{sec:CheMPS-ED-Comparison}

Figure~\ref{fig:rlm-coeffs-illustrate} contains the results of our
comparison of \CheMPS and ED moments for a fixed set of \CheMPS
parameters, stated in the figure legend.
Figures~\ref{fig:rlm-coeffs-illustrate}(a,b) show \cheby moments
$\mu_n$ and $\tilde \mu_n$, Figs.~\ref{fig:rlm-coeffs-illustrate}(c,d)
the $n$-dependent error measures, $\delta^\ED_n$, $\tilde
\delta^\ED_n$ and $ \tilde \delta^\Ch_n$.  From
Fig.~\ref{fig:rlm-coeffs-illustrate}(c) we note several points: (i)
For $n \le \Nmax$, the $\mu_n$-moments from \CheMPS and ED agree to
within about 1\%; this illustrates that \CheMPS is able to generate
rather accurate results for several hundered moments at modest
computational costs. (ii) $\mu_n$-moments are more accurate than
$\tilde \mu_n$-moments; the reason is that each $\mu_n$-moment depends
on only one \cheby vector, whereas each $\tilde \mu_n$-moment depends
on two. (Note, though, that if spectral reconstruction is performed by
employing both $\mu_n$-moments for $n \le \Nmax$ and $\tilde
\mu_n$-moments for $n> \Nmax$ (as done, e.g., for
Figs.~\ref{fig:finite-size-analysis} and \ref{fig:rmlm-all}), the
reduced accuracy of the $\tilde \mu_n$-moments is offset to some
extent if damping factors $g_n$ are employed, since these decay to 0
as $n$ approaches $N$, see inset of \fref{fig:kernel-delta}.) (iii)
The error measures $ \tilde \delta^\Ch_n$ and $\tilde \delta^\ED_n$
are of comparable magnitude; this implies that $ \tilde \delta^\Ch_n$
is a useful error quantifyer if exact results are not available.

The way in which theses errors depend on the various \CheMPS control
parameters can conveniently be analysed using the cumulative error
measures $ \Delta^{\ED}$, $\tilde \Delta^{\ED}$, $\tilde \Delta^\Ch$
and $\Delta^A$. These are shown in
Figs.~\ref{fig:rlm-coeffs-illustrate}(e,f) for various combinations of
$m$, $\ntrunc$ and $\dtrunc$. Several observations can be made: (iv)
When increasing the Krylov subspace dimension $\dtrunc$, all
cumulative errors decrease from $\dtrunc = 20$ to 30, but the decrease
saturates beyond $\dtrunc = 30$. (v) Increasing the number of energy
truncation sweeps beyond $\ntrunc = 10$ does not necessarily reduce
the cumulative errors; on the contrary, most actually increase,
implying that energy truncation sweeping should not be overdone.
(iv) The cumulative errors depend only weakly on the MPS dimension
$m$ (except for $\dtrunc = 10$, which is unreliable anyway), and
tend to be smaller(!) for $m=32$ than 64 (compare
Figs.~\ref{fig:rlm-coeffs-illustrate}(e) and
\ref{fig:rlm-coeffs-illustrate}(f)). This trend suggest that the
errors introduced by energy truncation grow if the mismatch between
$m$ and $\dtrunc$ grows.  Points (iv) to (vi) indicate that energy
truncation is the limiting factor for reducing \CheMPS errors, a
fact that will be elaborated on in
Sec.~\ref{sec:errors-recurrence-truncation} below.

To identify an optimal combination of \CheMPS control
parameters, we have collected error data such as those shown 
in Figs.~\ref{fig:rlm-coeffs-illustrate}(e,f) for
each possible combination of $\Wast=(1.1, 1.5, 2.0)$, $\epsilont=(0.1,
0.01, 0.025)$, $\dtrunc=(10, 20, 30, 50)$, $\ntrunc=(5, 10, 20)$, and
several $m$-values, for fixed maximum recursion number $\Nmax = 50$
and convergence threshold $\Delta_{\rm c} = 10^{-8}$. We concluded
that the choices $\dtrunc = 30$, $\ntrunc = 10$, $\Wast = 2 \WA$ and
$\epsilont = 0.025$ robustly yield good results (also for the HAFM),
and hence list these as recommended values in
Table~\ref{tab:expansion-params}.  Actually, the precise choice of
$\epsilont$ has only small effects on the error, as long as $\Wast$ is
chosen big enough. If $\Wast$ is too small, however,  the resulting spectral
function will loose some weight at high frequencies, because numerical
errors may cause energy truncation to effectively also project out
some contributions with energies smaller than the energy truncation
threshold $\etrunc$.

\subsection{Errors induced by recursion fitting and energy truncation}
\label{sec:errors-recurrence-truncation}

To better understand the error dependence on $m$, $\dtrunc$ and
$\ntrunc$ observed in points (iv) to (vi) of
Sec.~\ref{sec:CheMPS-ED-Comparison} above, let us analyse in more
detail the errors generatured during recurrence fitting
(Sec.~\ref{sec:cheby-iteration-step}) and energy truncation
(Sec.~\ref{sec:dmrg-energy-trunc}).
The error incurred when constructing $|t_n \rangle$ from $|t_{n-1}
\rangle$ and $|t_{n-2} \rangle$ using recurrence fitting is
characterized by the relative fitting error $\Delta^{\rm r}_{\rm fit}
= {\Delta_{\rm fit}}/{\norm{\ket{t_{n}}}}$ (\Eq{eq:fitting-norm}). The
effect of projecting out high-energy states using energy truncation,
$\ket{t_{n}} \mapsto P_\trunc \ket{t_n}$, can be characterized by the
average truncated weight per site during one truncation sweep,
$N_{\trunc}^{\sweep}$ (\Eq{eq:trunc-norm-per-sweep}), and by the
relative truncation-induced state change $\Delta^{\rm r}_{\trunc} =
{\Delta_{\trunc}}/{\norm{\ket{t_{n}}}}$ (\Eq{eq:Delta_trunc}).  The
latter measures intended changes in the state due to the truncation of
high energy weight, but also incorporates the effects of unavoidable
numerical errors.
\begin{figure*}[tb]
  \centering
  \includegraphics[width=\linewidth]{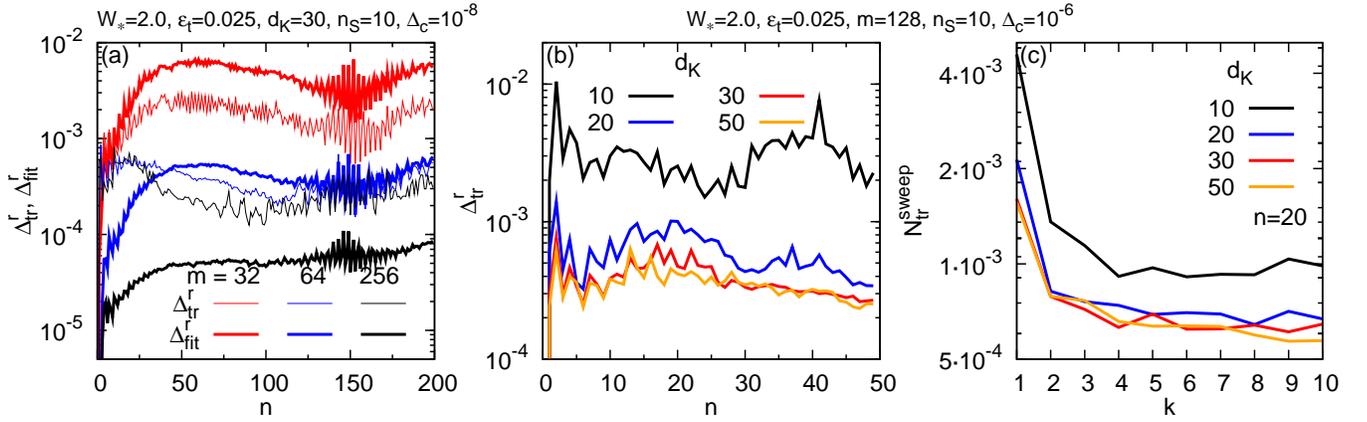}
  \caption{(Color online) (a) Relative fitting error $\Delta^{\rm
      r}_{\rm fit} = \Delta_{\rm fit}/\norm{\ket{t_{n}}}$
    (\Eq{eq:fitting-norm}) and relative truncation-induced state
    change $\Delta^{\rm r}_{\trunc} =
    \Delta_{\trunc}/\norm{\ket{t_{n}}}$ (\Eq{eq:Delta_trunc}), as
    functions of recursion number $n$, for three different choices of
    MPS dimension $m$.  Both quantities decrease with increasing $m$,
    but $\Delta^{\rm r}_{\rm fit}$ more strongly so since
    recurrence fitting is a strictly variational procedure, 
    whereas energy truncation is not.  (b)
    $\Delta^{\rm r}_{\trunc}$ as function of $n$, and (c) the average
    truncated weight per site $N_{\trunc}^{\rm sweep}$ (for $n=20$) as
    function of truncation sweep number $k$. Both (b) and (c) show
    results for four choices of Krylov subspace dimension $\dtrunc$,
    whose $\dtrunc$-dependence saturates beyond $\dtrunc = 30$.  }
  \label{fig:fidelity-compare}
\end{figure*}

These quantities are analysed in \fref{fig:fidelity-compare} in
dependence on $m$, $\dtrunc$ and $\ntrunc$. Continuing our list of
observations from the previous subsection, we note the following
features: (vii) Both $\Delta^{\rm r}_{\rm fit}$ and $\Delta^{\rm
  r}_{\trunc} $ are smaller than 1\% already for $m=32$
(\fref{fig:fidelity-compare}(a)), in accord with similar error margins
for $\delta^\ED_n$ in \fref{fig:rlm-coeffs-illustrate}(c).  (viii)
Both $\Delta^{\rm r}_{\rm fit}$ and $\Delta^{\rm r}_{\trunc} $
decrease with increasing $m$, but $\Delta^{\rm r}_{\trunc} $ does so
more slowly, and its decrease seems to saturate beyond $m=64$.  This
implies that \emph{energy truncation is the main limiting factor} for
\CheMPS.  The reason is that the intended purpose of energy
truncation, namely to strip $|t_n \rangle$ from its high-energy
components, modifies it in a way whose errors cannot be reduced to
arbitrarily small values. Indeed, this is illustrated by the following
two points: (ix) While both $\Delta_{\trunc}$ and $N_{\trunc}^{\rm
  sweep}$ initially decrease with increasing Krylov subspace dimension
$\dtrunc$, the decrease saturates for $\dtrunc \gtrsim 30$
(\fref{fig:fidelity-compare}(b,c)); (x) While $N_{\trunc}^{\rm sweep}$
initially decreases with the number of sweeps $\ntrunc$, the decrease
saturates already for $\ntrunc \lesssim 10$
\fref{fig:fidelity-compare}(c). Qualitatively, the behavior
shown in \fref{fig:fidelity-compare}(c) is robust. (However, the
choices of other \CheMPS control parameters do influence its
quantitative details, such as the $\dtrunc$ beyond which
$N_{\trunc}^{\rm sweep}$ becomes $\dtrunc$-independent.)  The lack
of saturation of $N_{\trunc}^{\rm sweep}$ with $\ntrunc$ implies
that there is no automatic stopping criterion for truncation
sweeps. Instead, the choice of $\ntrunc$ can be optimized as
described in Sec.~\ref{sec:CheMPS-ED-Comparison}, where 
we already concluded that taking $\ntrunc$ much larger than 10 
actually deteriorates the results.

Of course, truncation-induced errors can be avoided by simply
using the full bandwidth, $\Wast = W$, for which no trunctation is
necessary. However, in our experience the gain in resolution
obtained by using, instead, an effective bandwidth $\Wast \ll W$,
outweighs the small loss in accuracy incurred by 
the necessity to then perform energy truncation.

\section{Density matrix spectra}
\label{sec:density-matrix}

\begin{figure}[tb]
  \centering
  \includegraphics[width=\linewidth]{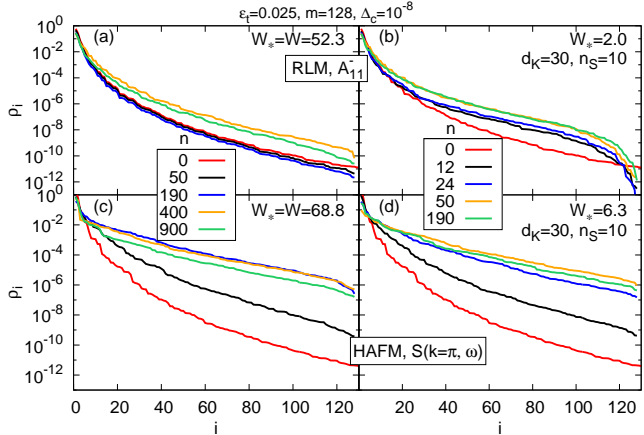}
  \caption{(Color online) Eigenvalue spectra $\rho_n(i)$ of the
    reduced density matrix at the center of the system for several
    expansion vectors $\ket{t_{n}}$ of (a,b) the RLM with $L_b = 101$,
    and (c,d) the HAFM with $L=100$. In (a,c) we used the full
    many-body bandwidth $\Wast = W$ without energy truncation,
    in (b,d) a reduced effective bandwidth with energy 
    truncation.}
  \label{fig:dm-spectrum}
\end{figure}

\begin{figure}[tb]
  \centering
  \includegraphics[width=\linewidth]{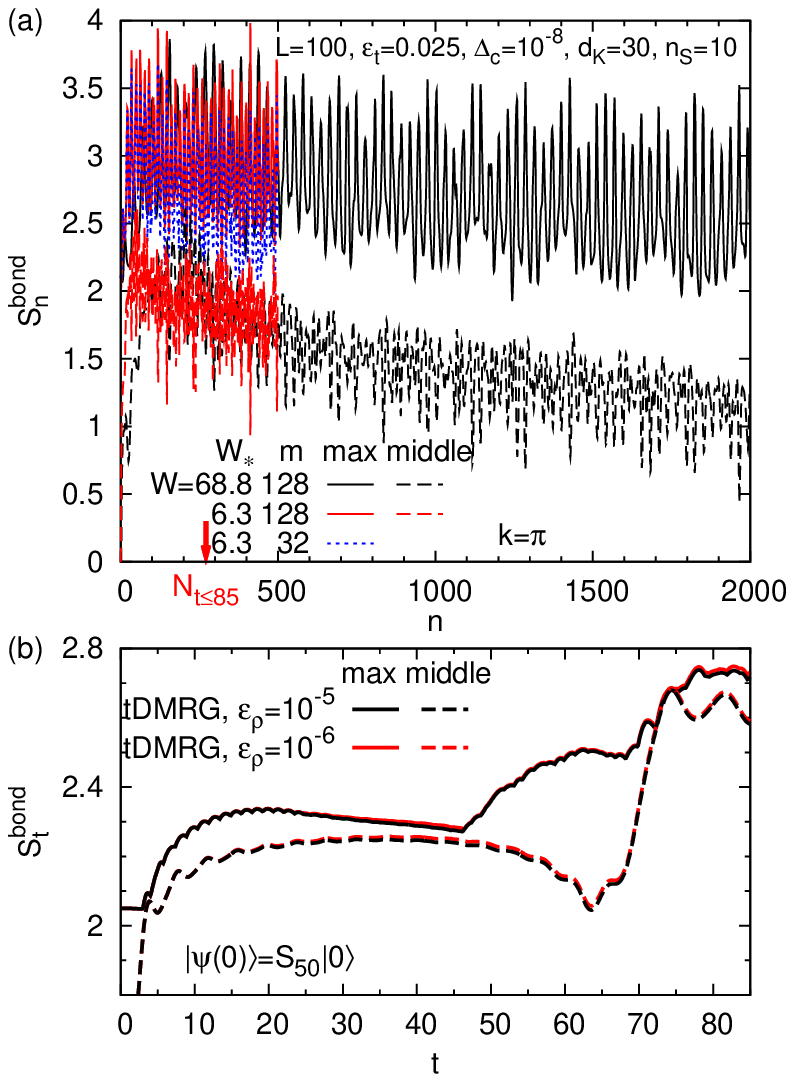}
  \caption{(Color online) Entanglement or bond entropy $S^{\rm
      bond}$ for the $k=\pi$ spectral function of the HAFM.  (a)
    $S^{\rm bond}_n$ for the \cheby vectors $\ket{t_{n}}$ and (b)
    $S^{\rm bond}_t$ during the tDMRG time evolution of $e^{-i \hat Ht } \hat
    {\vec{S}}_{x=50}\ket{0}$. In both (a) and (b), solid and dashed lines show
    the maximum bond entropy and the bond entropy at the middle of the
    system, respectively. (a) $S^{\rm bond}_n$ is shown for two
    choices of $\Wast$; the dotted line is from a calculation with a
    reduced $m=32$ and some entropy is lost due to truncation. The red
    arrow marks the expansion order roughly necessary to reach the
    time $t=85$ using the t\CheMPS technique for $\Wast=6.3$, here
    $N_{t\le 85}=271$. To reach the same time using $\Wast=W=68.8$ an
    order of expansion of $N_{t\le 85}=2961$ would be necessary. (b)
    $S^{\rm bond}_t$ is shown for two choices of the truncation
    error $\epsilon_\rho$. }
  \label{fig:entropy}
\end{figure}

The effects of energy truncation can be understood in more detail
by considering the reduced density matrix
\begin{eqnarray}
  \label{eq:reducedDM}
  \hat \rho_n  = {\rm Tr}_{\rm half} |t_n \rangle \langle t_n | 
\end{eqnarray}
where the trace is over one half of the chain. Let us analyse the
$n$-dependence of the spectrum of its eige<nvalues, say $\rho_n (i)$.
It can be used to quantify the entanglement encoded in $|t_n \rangle$,
via the associated \emph{entanglement} or \emph{bond entropy}, 
\begin{eqnarray}
  \label{eq:bond-entropy}
S_n^{\rm bond} = - \sum_i
\rho_n (i) \ln (\rho_n (i)) \; .   
\end{eqnarray}
Figure~\ref{fig:dm-spectrum} shows such density matrix 
spectra for both the RLM (panels (a,b)) and the HAFM (panels (c,d)),
calculated using both the full many-body bandwidth $\Wast = \Wfull$
(panels (a,c)) and a smaller effective bandwidth $\Wast$ (panels
(b,d)).  The $n=0$ line in all panels shows the eigenvalue spectrum
$\rho_0 (i)$, which reflects the entanglement encoded in $\ket{t_{0}}
= \Cop \ket{0}$ at the start of the recursion procedure.  In principle
one would expect the entire spectrum of density matrix eigenvalues
$\rho_n(i)$ to shift or rise to higher values as $n$ increases, since
multiplying $|t_{n-1} \rangle $ by $\hat H'$ when calculating $|t_n
\rangle$ (cf. \Eq{eq:dmrg-recipe-3}) generates entanglement entropy.
Such a \emph{spectral rise} with increasing $n$ is indeed observed in
all four panels of \ref{fig:dm-spectrum}, but the rise eventually
saturates for sufficiently large $n$. The speed of the initial stages
of the rise differs from panel to panel.  For the density matrix
spectra calculated \emph{without} energy truncation
(\fref{fig:dm-spectrum}(a,c)), the initial rise is rather slow, in
particular for the RLM (\fref{fig:dm-spectrum}(a), where the rise is
preceded by a slight initial decrease), reflecting the lack of strong
correlations of this model. In contrast, for density matrix spectra
calculated \emph{with} energy truncation
(\fref{fig:dm-spectrum}(b,d)), the initial rise is very rapid, and its
subsequent saturation sets in at quite small $n$ (of order 20 to 30).
Thus, energy truncation evidently has the effect of increasing
entanglement entropy. The reason is that the latter is calculated in a
different basis (the eigenbasis of $\hat \rho_n$) than that used to
perform energy truncation (the local eigenbasis of $\hat H'$).

According to \fref{fig:dm-spectrum}(d), the small MPS dimension of
$m=32$ used for the HAFM in \fref{fig:finite-size-analysis}(a) in
effect amounts to discarding the contributions to the reduced density
matrix of all states with weight below a threshold of around
$10^{-3}$. This threshold is rather large compared to typical DMRG
calculations, where characteristic truncation errors lie in the range
$10^{-6}$ to $10^{-8}$. It is remarkable that \CheMPS is nevertheless
able to give rather accurate results (such as reproducing CV results
obtained using $m_{\rm CV} =1500$). 

This efficiency appears to be an intrinsic feature of \CheMPS,
arising from the recursive manner in which the \cheby vectors $|t_n
\rangle$ are constructed. Evidence for this conclusion is presented
in \fref{fig:entropy}(a), which shows the bond entropy $S_n^{\rm
  bond}$ associated with $|t_n \rangle$ as function of recursion
number $n$. Remarkably, the bond entropy shows no tendencies towards
unbounded growth, even up to values as large as $n= 2000$. Quite
to the contrary: although the bond entropy increases somewhat when
increasing $m$ from 32 to 128 (with $\Wast = 6.3$), for either
case it tends to \emph{decrease} with recursion number $n$, and
similarly for the choice $\Wast = W$ without energy truncation.  All
of this is very encouraging, since it indicates that $n$ can be
increased, apparently at will, \emph{without incurring any runaway
  growth of DMRG truncation errors}. The reasons for this fact will be
recapitulated in the summary below.

For comparison, \fref{fig:entropy}(b) shows the bond entropy $S^{\rm
  bond}_t$ of a tDMRG calculation of the time evolution of
$|\psi(t)\rangle = e^{-i \hat Ht} \hat {\vec S}_{x = 50}|0\rangle$.
This entropy is, overall, smaller than the $S^{\rm bond}_n$ of the
\cheby vectors, because the initial state for the time evolution
involves an excitation at only one site, whereas the starting state
for the \CheMPS recursion involved a linear combination of local
excitations, $\hat {\vec{S}}_k|0\rangle$ (see \Eq{eq:spin-k}). The
most striking difference between $S^{\rm bond}_n$ and $S^{\rm
  bond}_t$, however, is that the former shows no trend to increase
with $n$, whereas the latter does with $t$. The increase in $S^{\rm
  bond}_t$ occurs in spurts, that happen each time a spin wave gets
reflected from one of the ends of the system, at which point more
numerical resources are required to keep track of the superposition of
incident and reflected spin waves.  For the present problem, the
increase in $S^{\rm bond}_t$ was not severe and remained completely
under control (staying below $S_n^{\rm bond}$
throughout). Nevertheless, we do believe that the contrast between
\fref{fig:entropy}(a) and \fref{fig:entropy}(b), showing a
nonincreasing trend for $S_n^{\rm bond}$ vs. an increasing trend for
$S_t^{\rm bond}$, is striking and significant.  It suggests that for
situations that feature strong entanglement growth with time, t\CheMPS
might be a promising alternative to tDMRG.

\section{Summary}
\label{sec:conclusion}

In this work, we have described \CheMPS as a method for
calculating zero-temperature spectral functions of one-dimensional
quantum lattice models using a combination of a \cheby expansion and
MPS technology.  To summarize our analysis, we would like to highlight what
we believe to be the two most important features of \CheMPS,
namely its efficiency and its control of spectral resolution.

\emph{Efficiency.} The first main feature is that \CheMPS provides an
\emph{attractive compromise between accuracy and efficiency}. It is
capable of reproducing correction vector results in the frequency
domain and tDMRG results in the time domain with comparably modest
numerical resources. In particular, surpringly small values for the
MPS dimension of $m$ are sufficient, even for obtaining spectral
resolution high enough to resolve finite size effects in great
detail. (For example, $m=32$ sufficed for the spin-$\frac{1}{2}$
antiferromagnetic Heisenberg model.) This remarkable efficiency, which
we had not anticipated when commencing this study, appears to be a
consequence of several factors: (i) \CheMPS does not suffer from a
runaway growth of DMRG truncation error with increasing $n$, because
the information needed to construct the spectral function with a
specified accuracy, say $\mathcal{O}(1/N)$, is not encoded in a single
state, but uniformly distributed over $N$ distinct \cheby vectors
$|t_n \rangle$.  (ii) These can be determined from Chebychev
recurrence relations involving only three terms, so that it is never
necessary to accurately represent the sum of more than two MPS.  (iii)
Moreover, these recurrence relations are numerically stable, i.e.\ the
inaccuracies in the calculation of \cheby vectors $|t_n \rangle$ do
not cause the \cheby expansion to diverge. (iv) Finally, the accuracy
needed for each $|t_n \rangle$ is set by that needed for $\mu_n =
\bra{0}{\cal B}\ket{t_{n}}$ (\fref{eq:dmrg-recipe-1}), which does not
need to be better than the specified accuracy, namely $\mathcal{O}
(1/N)$.

For spectral functions with a finite spectral width $\WA$ (which
is typically much smaller than the many-body bandwidth $W$), \CheMPS
offers a further attractive feature for enhancing efficiency: one
may use an ``effective bandwidth'' $\Wast$ of order $\WA$ (we
typically take $\Wast = 2 \WA$), which enhances spectral resolution
by a factor $W/\Wast$, at the cost of requiring additional energy
trunctation sweeps.  The latter are not necessary if one takes
$\Wast = W$, but then considerably higher expansion orders are
necessary to achieve comparable resolution.  In our experience the
benefits of enhanced resolution offered by the choice $\Wast = 2
\WA$ outweigh the costs of energy truncation.

\emph{Control of spectral resolution.}  The second main feature of
\CheMPS is that it offers \emph{very convenient control of the
  accuracy and resolution} of the resulting spectral function, by
simply adjusting the expansion order $N$. This is particularly useful
for studying finite-size effects, as exemplified in
\fref{fig:finite-size-analysis}.  On the one hand,
\fref{fig:finite-size-analysis}(b) shows very strikingly that the
structure factor an HAFM chain of finite length is dominated by a set
of discrete subpeaks which may be associated with the quantized
eigenenergies of spin wave excitations in a finite system. \CheMPS
allows the energies and weights of these excitations, and their
dependence on $L$, to be determined with unprecedented accuracy and
ease, by simply increasing $N$ until the peaks are well resolved.  On
the other hand, \fref{fig:finite-size-analysis}(f) shows that the
limit $L \to \infty$ may be mimicked by choosing $N$ just small enough
that the finite-size subpeaks are smeared out. Though the peak
shape thus obtained is slightly overbroadened (see inset of
\fref{fig:finite-size-analysis}(f)), this overbroadening can be
eliminated completely (see \fref{fig:discrete-spectrum}) by using a
discrete representation of the spectral function, that uses the
energies and weights of the discrete subpeaks as input. The ability
to fully eliminate overbroadening effects even for very large
many-body systems is, to the best of our knowledge, a unique feature
of \CheMPS.

On a technical level, the implementation of \CheMPS requires only
standard MPS techniques, such as the addition of different states and
the multiplication of operators. For energy truncation, single-site
sweeping needs to be set up with a new kind of local update, as
described in \fref{sec:dmrg-energy-trunc}. However, this procedure is
not too different from other known local update prescriptions and can
be implemented with modest programming effort.

\section{Outlook}
\label{sec:outlook}

Regarding future applications of \CheMPS, two directions for
further methodological development appear particularly
promising, namely time-dependence and finite temperature. A few 
comments are due about each.

\emph{Time dependence.} While the good agreement between t\CheMPS and
tDMRG reported in \fref{fig:time-skt} is encouraging, a detailed
analysis of t\CheMPS should be performed to understand the nature of
its error growth with time, and to explore under which conditions, if
any, t\CheMPS offers competitive advantages relative to tDMRG.  On the
one hand, tDMRG has the advantage that highly efficient Krylov methods
can be used to optimize the evaluation of $e^{-i \hat H \Delta t}
|\psi (t)\rangle $ w.r.t.\ the state $|\psi (t)\rangle $ being
propagated; however its numerical costs increase rapidly if $|\psi (t)
\rangle$ contains a broad spectrum of excited states.
On the other hand, \CheMPS has the advantage (i) that the Chebychev
expansion of the operator $e^{-i \hat H t}$ can be applied with equal
accuracy to every state in the Hilbert space, in particular also
highly excited ones.  Moreover, (ii) very large evolution times might
be achieved more easily with t\CheMPS than tDMRG, since the former
represents $|\psi(t)\rangle$ as a sum over many \cheby vectors (see
\Eq{eq:lin-comb}), thereby being potentially less susceptible than
tDMRG to the growth of truncation errors (as discussed in the
introduction, and exemplified in \fref{fig:entropy}). We expect that
for some applications (i) and/or (ii) may offer advantages for
t\CheMPS over tDMRG, e.g.\ for calculating quantum quenches starting
from strongly nonequilibrium initial states, but leave a detailed
investigation to the future.

\emph{Finite temperature.}  The fact that \CheMPS uniformly
resolves the entire energy spectrum of $\hat H$ suggests that it should
be particularly suited for calculating the spectral functions
$\mathcal{A}_T^{\mathcal{BC}} (\omega) = \int \frac{d t}{2 \pi}e^{i \omega t} G_T^{\mathcal{BC}} (t)$
of finite temperature correlators such as
\begin{eqnarray}
  \label{eq:GAB-finite-T}
  G_T^{\mathcal{BC}} (t) = {\rm Tr} [\hat \rho_T \Bop (t) \Cop(0) ] \; , \quad \hat \rho_T = \frac{e^{-\beta \hat H} }{Z} .
\end{eqnarray}
According to Ref.~\onlinecite{Weisse2006}, such a spectral function can be evaluated using
\cheby expansions by proceeding as follows: Express the partition function as 
\begin{subequations}
  \begin{eqnarray}
    \label{eq:partition-function}
    Z =  \int d \omega e^{-\beta \omega} \rho(\omega) \; , 
  \end{eqnarray}
  by introducting the density of states 
  \begin{eqnarray}
    \label{eq:DOS}
    \rho (\omega) = {\rm Tr}[\delta (\omega - \hat H)] \; ,
  \end{eqnarray}
\end{subequations}
and the spectral function as 
\begin{subequations}
  \begin{eqnarray}
    \label{eq:spectral-frequency-finite-T}
    \mathcal{A}_T^{\mathcal{BC}}(\omega) & = & \frac{1}{Z} \int d \bar \omega \, e^{-\beta \bar \omega} \mathcal{\rho}^{\mathcal{BC}} 
    (\bar \omega, \omega + \bar \omega) \; , 
  \end{eqnarray}
  by introducing the  \emph{density of matrix elements}\cite{Wang1994,Wang1994a} 
\begin{eqnarray}
  \label{eq:BC-matrix-element-density}
  \mathcal{\rho}^{\mathcal{BC}}(\bar \omega , \omega) & = & {\rm Tr} [ \delta(\bar \omega - \hat H ) \, \Bop \, \delta(\omega - \hat H ) \, \Cop ] \; .
\end{eqnarray}
\end{subequations}
Then \cheby expand the $\delta$-functions in
\Eqs{eq:DOS} and (\ref{eq:BC-matrix-element-density}) using
\Eq{eq:spectral-exp-A} (after suitably rescaling Hamiltonian and
frequencies).  The resulting \cheby expansions will contain moments of
the form
\begin{subequations}
  \label{sec:T-moments}
  \begin{eqnarray}
    \mu_n^\rho & = & {\rm Tr}[T_n (\hat H')] \; , \\
    \mu_{nn'}^{\mathcal{BC}} & = & {\rm Tr}[T_{n}(\hat H') \, \Bop \, T_{n'}(\hat H') \, \Cop] \; .
  \end{eqnarray}
\end{subequations}
We now note that this framework is very well suited for an MPO
implementation, which would consist of three steps: (i) Using \cheby
recurrence relations, recursively construct and store MPO
representations for each operator $T_{n}(\hat H')$; we expect
(based on our experience with the \cheby vectors $|t_n \rangle$)
that this should be possible without runaway costs in numerical
resources, since the construction of $T_{n}(\hat H')$ requires only
$\hat H' T_{n-1}(\hat H')$ and $T_{n-2}(\hat H')$. (ii) Calculate
the moments in \Eqs{sec:T-moments} by evaluating the traces,
which is straightforward in the context of MPS/MPO. (iii)
Insert the resulting moments into the reconstructed Chebychev
expansions for $\rho(\omega)$ and $\rho^{\mathcal{BC}}(\bar \omega,
\omega)$, and finally evaluate the integrals
\Eqs{eq:partition-function}) and
(\ref{eq:spectral-frequency-finite-T}).  Note the economy of this
scheme: after once constructing the MPO for each $T_{n}(\hat H')$, and
once evaluating the trace for each moment $\mu_n^\rho$ and
$\mu_{nn'}^{\mathcal{BC}}$, the spectral function
$\mathcal{A}_T^{\mathcal{BC}}(\omega) $ can be calculated for
arbitrary combinations of $\omega$ and $T$. The implementation of this
strategy is left for future studies.

We conclude by remarking that the idea of using \cheby expansions in
the context of many-body numerics, advocated in inspiring fashion in
Ref.~\onlinecite{Weisse2006}, can be implemented in combination with
any method that is able to efficiently apply a Hamiltonian $\hat H$ to
a state $|\psi \rangle$. \cheby expansions optimize the resolution
that can be extracted from a limited number of applications of $\hat
H$. While \CheMPS is based on doing this using MPS methods for
one-dimensional lattice models, similar developments have been
pursued within the context of exact
diagonalization\cite{Alvermann2007,AlvermannFehske2009} and Monte
Carlo\cite{Weisse2009} methods, and \cheby expansions should also be
useful in combination with tensor network methods for
two-dimensional quantum lattice models.

\begin{acknowledgments}
  We thank A. Wei{\ss}e for an inspiring talk on kernel polynomial
  methods, which motivated us to implement the ideas of
  Ref.~\onlinecite{Weisse2006} using MPS technology; T.~Barthel and
  J.-S. Caux for providing the tDMRG and Bethe Ansatz data,
  respectively, that are shown in Figs.~\ref{fig:finite-size-analysis}
  to \ref{fig:discrete-spectrum}; and J. Halimeh for help with
  extracting the discrete data shown in
  Fig.~\fref{fig:discrete-spectrum} from large-$N$ \CheMPS spectra.
  We gratefully acknowledge helpful discussions with P. Schmitteckert,
  who independently pursued ideas similar to those presented here,
  A. Alvermann, T. Barthel, H. Fehske and M. Vojta.  This work was
  supported by DFG (SFB 631, De-730/3-2, SFB-TR12, SPP 1285,
  De-730/4-1). Financial support by the Excellence Cluster
  ``Nanosystems Initiative Munich (NIM)'' is gratefully acknowledged.
\end{acknowledgments}

\appendix

\section{Resonant level model}
\label{sec:resonant-level-model}

This appendix introduces the fermionic resonant level model (RLM)
that was used for the error analysis of
Sec.~\ref{sec:error-analysis}, and presents \CheMPS results for its
spectral functions.

The RLM is defined by the following Hamiltonian:
\begin{eqnarray}
  \label{eq:rlm-ham-normal}
  \hat H_{\rm RLM} &= & \sum_{i=1}^{n_{d}} \varepsilon_{i} \hat d{}_{i}^{\dagger} \hat d{}_{i}^{\phantom{+}} + 
  \sum_{i=1}^{n_{d}} V_{i} \sum_{k=1}^{L_{b}} \left(\hat d{}_{i}^{\dagger} \hat c{}_{k}^{\phantom{+}} + \hc\right)\\
  \nonumber
  & & + \sum_{k=1}^{L_{b}}\varepsilon_{k} \hat c{}_{k}^{\dagger} \hat c{}_{k}^{\phantom{+}} \; . 
\end{eqnarray}
It describes a set of $n_{d}$ discrete, ``local'', non-interacting fermion
levels with energies $\varepsilon_i$, that hybridize with strengths $V_i$
with a band of $L_b$ $(\gg 1)$ fermion levels with energies
$\varepsilon_k$, assumed uniformly spaced within the interval $[-W_b,
W_b]$. We choose $W_b = 1$ as unit of energy throughout this
section. We will parametrize the hybridization strengths $V_{i}$ in terms of the
associated level widths $\Gamma_{i} = \pi\frac{L_{b}}{2} V_{i}^{2}$.

The spectral function $A_{ij}(\omega) \equiv A_{ij}^+ (\omega) + A_{ij}^- (\omega)$ has two contributions,
\begin{equation}
  \label{eq:rlm-A-sum}
  A_{ij}^+ (\omega) \equiv \mathcal{A}^{d_{i}^{\phantom{*}}d_{j}^{\dagger}}(\omega), \qquad 
  A_{ij}^- (\omega) \equiv  \mathcal{A}^{d_{j}^{\dagger}d_{i}^{\phantom{+}}}(-\omega) \; , 
\end{equation}
describing particle and hole excitations, which at $T=0$ are
proportional to step functions $\theta (\pm \omega)$ that vanish for
$\omega < $ or $> 0$, respectively.  Since the RLM \ham is
quadratic, the problem can be solved by diagonalizing 
the single-particle problem. In the continuum limit 
$L_{b} \rightarrow\infty$, this yields the following
exact expression for the spectral function,\cite{Hewson1997}
for $|\omega| < D_b = 1$:
\begin{eqnarray}
  \label{eq:rlm-A-cont}
  A^{\infty}_{ij}(\omega) & = & \lim_{\eta\rightarrow 0^{+}}-\frac{1}{\pi} \Im \left(\left[\omega + i\eta - \Upsilon - \Delta(\omega)\right]^{-1}\right)_{ij},\\ 
  \nonumber
  \Upsilon_{ij} & = & \varepsilon_{i}\delta_{ij} , \quad \Delta_{ij}(\omega) = \frac{1}{\pi}\sqrt{\Gamma_{i}\Gamma_{j}}
  \left( \ln \left| \frac{\omega-D_{b}}{\omega+D_{b}}\right| - i \pi \right) \;,
\end{eqnarray}
where $\Upsilon$ and $\Delta$  are matrices of dimension
$n_d \times n_d$.  The \cheby moments $\mu_{n}$ for
the finite system of length $L$ can also be found exactly, by
evaluating the expectation values \Eq{eq:spectral-exp-mu-res} using
the (numerically-determined) exact single-particle eigenstates of $\hat H$.
\begin{figure*}[tb]
  \centering
  \includegraphics[width=\linewidth]{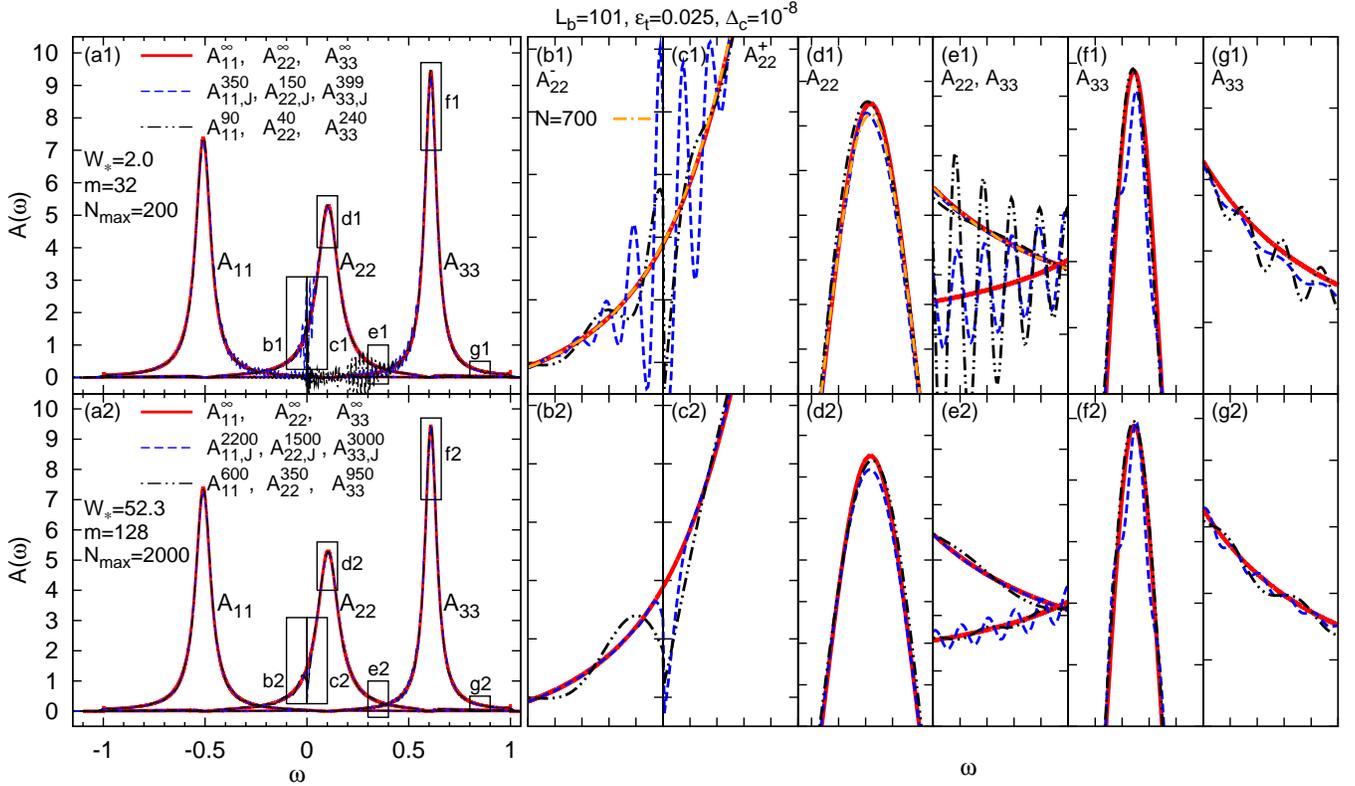}
  \caption{(Color online) Diagonal spectral functions
    $\mathcal{A}_{jj}(\omega)$ of a 3-level RLM. Thick solid lines
    show the continuum limit, $A^{\infty}_{jj}(\omega)$, from
    \fref{eq:rlm-A-cont}. Dashed and dashed-double dotted lines show
    \CheMPS results, $\mathcal{A}^N_{jj,J}$ or $\mathcal{A}^N_{jj}$,
    with or without Jackson broadening, respectively, calculated for
    $L_b=101$ band levels. For each spectrum, the effective broadening
    $\etaN$ was taken as large as possible without lowering the peak
    height significantly below that of $A^\infty_{jj}$. In (a1-g1) we
    used an effective bandwidth of $\Wast=2.0$ and in (a2-g2) the full
    many-body bandwidth $\Wast= 52.3$. The latter requires
    significantly larger expansion orders, but exhibits less numerical
    inaccuracies, compare (b1-g1) and (b2-g2), which represent zooms
    of the rectangles indicated in (a1) and (a2), respectively.
    (b,c): Gibbs oscillations arise if $\mathcal{A}^\pm(\omega)$ are
    expanded separately, so that \CheMPS attempts to resolve their
    $\theta(\pm \omega)$ steps.  Expanding instead their sum,
    $\mathcal{A}^+ (\omega) + \mathcal{A}^- (-\omega)$, and performing
    a Jackson-damped reconstruction, we obtain the smooth
    dashed-dotted line in (b1-e1) (calculated using ED moments). (d,e)
    $\mathcal{A}^N_{22}$ nicely reproduces $ \mathcal{A}^\infty_{22}$,
    because the peak is somewhat broader than $\omega_L$.  (e-g) $
    \mathcal{A}^N_{33}$ shows small but distinct finite-size wiggles,
    because the main peak is so sharp and narrow that recovering its
    height fully, requires an $\etaN$ so small that it is comparable
    to $\omega_L$.}
  \label{fig:rmlm-all}
\end{figure*}

The Hamiltonian (\ref{eq:rlm-ham-normal}) corresponds to a ``star
geometry'', since each local level couples to every band level.  For
the purposes of using \CheMPS, however, it needs to be  transformed
to a ``chain geometry'' of the form
\begin{align}
  \label{eq:rlm-ham-tri}
  \begin{split}
    \hat H_{\rm RLM} &= \sum_{i=1}^{n_{d}} \varepsilon_{i}
    \hat d{}_{i}^{\dagger} \hat d{}_{i}^{\phantom{+}} +
    \sum_{i=1}^{n_{d}}\sqrt{\frac{2\Gamma_{i}}{\pi}}
    \left(\hat d{}_{i}^{\dagger} \hat f{}_{1}^{\phantom{+}} + \hc\right)\\
    &+ \sum_{\ell =1}^{L_{b}-1} \lambda_{\ell}
    \left(\hat f{}_{\ell}^{\dagger} \hat f{}_{\ell+1}^{\phantom{+}} + \hc\right).
  \end{split}
\end{align}
This can be achieved\cite{BullaCostiPruschke2008} using Lanczos
tridiagonalization of the band part of the \ham, thereby determining
the hopping coefficients $\lambda_{\ell}$.

Starting from \Eq{eq:rlm-ham-tri}, we have used \CheMPS to calculate
the diagonal components $\mathcal{A}_{jj}$ of the RLM spectral
function for a model with $n_d = 3$ local levels.  In contrast to
Section~\ref{sec:finite-size}, our interest here is not in analysing
finite-size effects, but in determining how the \CheMPS parameters
need to be adjusted to recover the exact continuum function
$\mathcal{A}_{jj}^\infty$ of \Eq{eq:rlm-A-cont}. Thus, we purposefully
chose a set of model parameters leading to three well-separated peaks
of slightly different widths, taking $\varepsilon_j \in \{-0.5, 0.1,
0.6\}$ and $\Gamma_j \in \{0.04, 0.06, 0.03 \}$, and chose the number
of band levels $L_b=101$ large enough that the finite-size spacing
$\omega_L \simeq 1/L_b = 0.01$ is somewhat smaller than the smallest
peak width, $\Gamma_3$.  By choosing the expansion order for each
curve such that the effective broadening lies in the window between
the finite-size spacing and the intrinsic peak width, $\omega_L <
\etaN < \Gamma_j$, it should be possible to reveal the shape of
$\mathcal{A}_{jj}^\infty$ quite accurately without yet resolving
finite-size subpeaks (though traces of the latter might show up for
$\mathcal{A}_{33}$, for which this window is small).  To this end, we
used the following criterion for choosing $N$ when reconstructing
$\mathcal{A}^N_{jj}$: the effective broadening $\etaN$ was taken as large as
possible without lowering the peak height significantly below that of
$A^\infty_{jj}$ (this corresponds to choosing $\etaN \lesssim \Gamma_j$). 

The results of these calculations are summarized in
Fig.~\ref{fig:rmlm-all}; all spectra shown there were obtained by
performing separate expansions for the positive and negative branches,
$A^\pm_{jj}(\omega)$ (with one exception, noted below).

Figures~\ref{fig:rmlm-all}(a1-g1) were calculated using an effective
bandwidth of $\Wast=2.0$ (with $\varepsilon_t=0.025$) for each branch,
corresponding to roughly twice the spectral width of each branch,
which is of order of the single-particle band-width, $W_A \simeq W_b =
1$.  For this choice, an MPS dimension of merely $m=32$ was found to
suffice for accurate recurrence fitting. Figure~\ref{fig:rmlm-all}(a1)
illustrates a number of points: (i) By choosing $\etaN$ according to
the above criterion of recovering the correct peak height, excellent
agreement with the continuum limit $A^\infty$ of \Eq{eq:rlm-A-cont} is
obtained over most of the frequency range.  (ii) This is the case both
with and without Jackson damping (thin black or blue lines,
respectively), but with Jackson damping, higher expansion orders are
needed to obtain the correct peak heights, since Jackson damping
induces some artificial broadening (by a factor of $\pi$, see
\Eq{eq:kernels-J}). (iii) Small oscillations remain in some frequency
ranges (see Figs.~\ref{fig:rmlm-all}(b1-g1) for zooms). These stem
from three sources: finite-size subpeaks, numerical inaccuracies and
step function artefacts near $\omega = 0$ (cf.\ points (iv), (vi) and
(viii) below, respectively).  (iv) For the spectrum with the narrowest
peak, $\mathcal{A}_{33}$, the window between $\omega_L$ and
$\Gamma_{33}$ is so small that the criterion of reproducing the
continuum peak height implies that small finite-size subpeak remain
visible, see Figs.~\ref{fig:rmlm-all}(e1-g1) for zooms. (v) In
contrast, such oscillations are almost entirely absent for the
broadest peak, $\mathcal{A}_{22}$ (see
Figs.~\ref{fig:rmlm-all}(d1,e1)), since its width $\Gamma_{2}$ is
somewhat larger than $\omega_L$.

In order to illustrate the effect of energy truncation,
Figs.~\ref{fig:rmlm-all}(a2-g2) show the same spectral functions as
\fref{fig:rmlm-all}(a1-g1), but now setting $\Wast=W$, the full
many-body bandwidth (here = 52.3), so that no energy truncation is
needed. This allows us to make some additional instructive
observations: (vi) Using the full bandwidth yields results of higher
quality, in that numerical artefacts are significantly weaker (except
near $\omega = 0$), compare Figs.~\ref{fig:rmlm-all}(d2-g2) and
(d1-g1). The reason is that energy truncation constitutes \CheMPS's
dominant source of error (as shown in Section~\ref{sec:error-analysis}
below); its avoidance thus yields more precise \cheby moments $\mu_n$,
especially for $n>\Nmax$.  (vii) However, this improvement is
numerically expensive: the increased effective bandwidth neccessitates
larger expansion orders $N$, which in turn requires a higher MPS
dimension (here $m=128$). (viii) For the present model, it was
possible to calculate several thousand moments without encountering
numerical instabilities; this illustrates the fact that the Chebychev
recurrence relations are \emph{numerically stable}.

Finally, let us address (ix) the wiggly artefacts near $\omega = 0$.
They reflect the fact that \CheMPS was \emph{separately} applied to
the positive and negative branches of the spectral function,
$A^{\pm}(\omega)$, shown in zooms in Figs.~\ref{fig:rmlm-all}(b) and
\fref{fig:rmlm-all}(c), respectively.  These are proportional to step
functions $\theta (\pm \omega)$, and hence abruptly dip to zero for
$\omega <0$ or $> 0$, respectively.  The wiggly artefacts correspond
Gibbs oscillations decorating these sharps dips.  This problem can be
avoided by performing a single \cheby expansion of the \emph{sum},
$\mathcal{A}^+ (\omega) + \mathcal{A}^- (-\omega)$, which is a smooth
function and leads to the perfectly smooth long-dashed line in
Figs.~\ref{fig:rmlm-all}(b,c).  This improvement comes at roughly
twice the numerical cost, since it requires a doubling of the spectral
range to $\omega \in [-W_A, W_A]$: this implies a slight but obvious
modification of the transformations from $\omega$ to $\omega'$ and
from $H$ to $\hat H'$ to account for the shifted range of $\omega$;
and a doubling of $W_\ast$ and hence of the expansion order $N$
required to achieve a specified resolution.

The main conclusions from our \CheMPS calculations for the RLM are as
follows: The strategy of using twice the spectral width as effective
bandwidth ($\Wast = 2 W_A$) and performing energy truncation
(\fref{fig:rmlm-all}(a)) is a satisfactory compromise between
efficiency (only a few hundred \cheby moments are needed) and accuracy
(for which energy truncation is the main limiting factor). If desired,
better results can be obtained by using the full bandwidth ($\Wast =
W$) and thus avoiding energy truncation, albeit at the cost of
significantly increasing the required expansion order by the factor
$W/2W_A $.  Nevertheless, the calculation of \cheby moments $\mu_n$
with very large $n$ is feasible due to the remarkably numerical
stability of Chebychev recurrence relations.

% \bibliography{refs}

\end{document}